\documentclass[12pt]{iopart}
\usepackage{iopams}  

\usepackage{psfrag,epsfig,amsfonts,amssymb,wasysym,bm}

\usepackage{bbold}
\usepackage[normalem]{ulem}
\usepackage{color}

\newcommand{\pr}{{\rm{Prob}}}

\newcommand{\id}{\mathbb 1}
\newcommand{\pu}{\mbox{Tr}\{[\rho(0)]^2\}}

\newcommand{\pmax}{p_{\!\!\;\rm{max}}}

\newcommand{\dia}{\!\!\;\rm{dia}}
\newcommand{\rhobar}{\rho_{\!\!\;\rm{dia}}}
\newcommand{\rhoaux}{\rho_{\!\!\;\rm{aux}}}

\newcommand{\Adia}{A^{\!\!\;\rm{dia}}}
\newcommand{\RR}{{\mathbb R}}

\newcommand{\tmax}{t_{\mathrm{max}}}

\newcommand{\kB}{k_\mathrm{B}}
\newcommand{\DD}{D}
\newcommand{\hh}{\hbar}

\newcommand{\propa}{{\cal U}}

\providecommand{\norm}[1]{\|#1\|}
\newcommand{\da}{\Delta_{\! A}}
\newcommand{\dda}{\delta \! A}

\begin{document}

\title{Transportless equilibration in isolated many-body quantum systems}

\author{Peter Reimann}
\address{Fakult\"at f\"ur Physik, 
Universit\"at Bielefeld, 
33615 Bielefeld, Germany}

\ead{reimann@physik.uni-bielefeld.de}

\begin{abstract}
A general analytical theory of temporal relaxation processes
in isolated quantum systems with many degrees of freedom
is elaborated, which unifies and substantially amends
several previous approximations.
Specifically, the Fourier transform of the initial 
energy distribution is found to play a key role, 
which is furthermore equivalent to the so-called 
survival probability in case of a pure initial state.
The main prerequisite is the absence of any notable 
transport currents,
caused for instance
by some initially unbalanced local densities
of particles, energy, and so on.
In particular, such a transportless relaxation
scenario naturally arises when both
the system Hamiltonian and the initial non-equilibrium
state do not exhibit any spatial inhomogeneities
on macroscopic scales.
A further requirement is that the relaxation
must not be notably 
influenced by
any approximate (but not exact) constant
of motion or metastable state.
The theoretical predictions are compared 
with various experimental and numerical 
results from the literature.
\end{abstract}

\maketitle

\vspace{2pc}

\section{Introduction and Overview}
\label{s0}
Relaxation processes in systems with many degrees of freedom 
play a key role in a large variety of different physical 
contexts \cite{mor18,gog16,ale16,tas16,bor16,nan15,gem09}.
Quite often, an essential feature of the pertinent 
non-equilibrium initial states are some unbalanced local 
densities of particles, energy, etc., giving rise to 
transport currents during the relaxation towards 
equilibrium.
Paradigmatic examples are compound systems, 
parts of which are initially hotter than others,
or a simple gas in a box, streaming through 
a little hole into an empty second box.
As a consequence, the temporal relaxation 
crucially depends on the system size, and may
become arbitrarily slow for sufficiently large 
systems.

In the present work, the focus is on the complementary
class of equilibration processes, which do not entail any 
such transport currents.
In the simplest case, one may think of systems without
any spatial inhomogeneities on the macroscopic scale,
for instance a fluid or solid with spatially constant 
densities of all particle species, energy, 
and so on.
(Inhomogeneities on 
the microscopic (atomic) scale are obviously still 
admitted; they are outside the realm to which 
concepts like ``densities'' and associated 
``transport currents'' are applicable, see also section \ref{s4}.)
The non-equilibrium character of an initial
state could then for instance manifest itself
in a non-thermal velocity distribution.
Another concrete experimental example, 
to which we will actually apply our 
theory in section \ref{s6}, is the excitation of
an ``electron gas'' by a laser pulse,
resulting in a system state, which is
spatially homogeneous but exhibits 
strong deviations from the usual 
Fermi-Dirac statistics at 
equilibrium.
Further pertinent examples, which are often
considered in numerical investigations, 
and which will also be compared with our
present theory later on, are so-called quantum
quenches,  where the initial state is given by
the ground state (or some other eigenstate or 
thermal equilibrium state) of 
a Hamiltonian, which is different from the Hamiltonian 
that governs the 
actual relaxation dynamics.
Still focusing on spatially homogeneous 
Hamiltonians and states, also other
types of ``handmade'' non-equilibrium
initial conditions are commonly explored
in the literature, e.g., so-called N\'eel 
states (antiferromagnetic order)
in the context of various spin models.
In all these cases of transportless equilibration, 
it is reasonable to expect (and will be confirmed 
later on) that the temporal relaxation is 
practically independent of the system size, 
and that the typical time scales will be much 
faster than for transport governed 
equilibration.
As yet another striking feature, we will find 
that transportless relaxation is usually 
not exponential in time.

The general issues of equilibration and 
thermalization in 
isolated many-body quantum 
systems have stimulated during recent 
years a steadily growing amount of analytical, 
numerical, as well as experimental activity, 
reviewed, e.g. in  
\cite{mor18,gog16,ale16,tas16,bor16,nan15,gem09}.
(In doing so, also open systems (interacting and 
possibly entangled with an environment) can be 
treated by considering the environment 
(thermal bath, particle reservoir etc.) and 
the actual system of interest as an isolated 
compound system.)
Strictly speaking, the relaxation of such an 
isolated system towards a steady long-time limit is
immediately ruled out by the unitary 
time evolution and, in particular, by the 
well-know quantum revival effects \cite{hob71}.
Nevertheless, ``practical equilibration''
(almost steady expectation values for
the vast majority of all sufficiently 
large times) has been rigorously 
established in  
\cite{equil,sho11,sho12,rei12,bal16}
under quite general conditions.

In section \ref{s1}, the essential points of 
those previous results on equilibration 
will be made plausible once again
by means of a new, less rigorous, 
but much simpler and intuitive 
reasoning.
It should be emphasized that the issue
of equilibration is related to, 
but different from the issue of
thermalization, i.e., the question 
whether or not the above mentioned (almost) 
steady expectation values in the 
long-time limit agree with the textbook
predictions of equilibrium 
statistical mechanics.
The latter issue of thermalization
does not play any role throughout
this paper: all results are valid
independently of whether or not the considered 
system thermalizes.

In section \ref{s2}, the previous rigorous
approach to transportless equilibration 
from  \cite{rei16,bal17} is revisited 
in terms of an alternative, non-rigorous 
but physically much simpler line of 
reasoning, while in sections \ref{s3} 
and \ref{s4} its main preconditions 
are worked out in considerable more 
detail than before.
A representative comparison of this
theory with experimental observations
is provided by section \ref{s6}.

Section \ref{s5} represents the actual 
core of the paper,
and the formal approach adopted in this 
section is substantially more 
elaborate
than in the previous sections 
\ref{s1} and \ref{s2}.
Technically speaking, the crucial idea
is to skillfully ``rearrange'' the 
systems's very dense energy eigenvalues 
and to ``redistribute'' the possibly 
quite heterogeneous populations of the 
corresponding eigenstates,
yielding an effective description 
in terms of an auxiliary 
Hamiltonian with approximately 
equally populated eigenstates.
The main result is a 
unification and substantial amendment
of the earlier findings 
in  \cite{rei16,bal17,tor14,tor15}, formally summarized by 
the compact final equation (\ref{600}).
The decisive quantity, which governs the
temporal relaxation via the last term 
in equation (\ref{600}), 
will furthermore be identified 
in section \ref{s5}
with the Fourier transform of the system's 
initial energy distribution, and
in case the system is in a pure state,
also with the so-called survival 
probability of the initial state.
These analytical predictions are 
compared with previously published 
numerical simulations in section \ref{s7}.

Even when focusing solely on analytical 
investigations, previous studies related 
to relaxation time scales and the like 
are still quite numerous, and pointing out 
in each case the similarities and differences 
to our present approach goes beyond 
the scope of this paper.
A first major issue in this context, 
addressed e.g. in  
\cite{sho12,bounds}, 
is the derivation of general
upper bounds for some suitably 
defined relaxation time.
While in some specifically tailored 
examples, the relaxation may indeed
become extremely slow 
\cite{slowfast}, 
those upper bounds are still not
quantitatively comparable
to the actually observed time 
scales in more realistic situations.
On the other hand, extremely fast time scales
have been predicted, e.g., in 
\cite{slowfast,fast}.
Finally, investigations of 
particular classes of models, 
observables, or initial conditions 
are provided, among others, 
in 
\cite{sre99,special}.
One important step forward of our 
present work is that not only an 
estimate of some characteristic time 
scale, but also a detailed 
description of the entire temporal 
relaxation behavior is provided
and quantitatively verified against 
experimental and numerical data.

\section{Equilibration and thermalization}
\label{s1}
We consider an isolated system, 
modeled by a Hamiltonian 
\begin{eqnarray}
H=\sum_n E_n\, |n\rangle \langle n|
\label{5}
\end{eqnarray}
and an 
initial state $\rho(0)$
(pure or mixed and in general far 
from equilibrium), which
evolves in time according to
\begin{eqnarray}
\rho(t)=\propa_t\rho(0)\propa_t^\dagger
\label{4}
\end{eqnarray}
with propagator 
\begin{eqnarray}
\propa_t:=e^{-iHt/\hh}
\ .
\label{6}
\end{eqnarray}
Hence, the expectation value
\begin{eqnarray}
\langle A\rangle_{\!\rho}:= \mbox{Tr}\{\rho A\}
\label{7}
\end{eqnarray}
of any given observable $A$ in the time evolved state
$\rho(t)$ follows as
\begin{eqnarray}
\langle A\rangle_{\!\rho(t)}
= \sum_{m,n} 
 e^{i[E_n-E_m]t/\hh} 
\,
\rho_{mn}(0) A_{nm}
\ ,
\label{10}
\end{eqnarray}
where 
$\rho_{mn}(t):=\langle m|\rho(t)|n\rangle$,
$A_{nm}:=\langle n |A| m \rangle$,
and where,
depending on the specific problem 
under consideration, 
the indices $n$ and $m$ run from $1$ 
to infinity or to some finite upper limit.
In particular, 
\begin{eqnarray}
p_n:=\langle n|\rho(0)|n\rangle=\rho_{nn}(0)
\label{11}
\end{eqnarray}
represents the
population of the $n$-th energy level,
i.e., the probability that the system is found
in the energy eigenstate $|n\rangle$ when
averaging over many repetitions of the
measurement and -- in the case of a mixed state --
over the statistical ensemble described
by $\rho(0)$.

The main examples we have in 
mind are macroscopic 
systems with, say, $f\approx 10^{23}$ 
degrees of freedom.
While such many-body quantum systems 
are generically non-integrable,
so-called integrable systems
are still admitted in most
of what follows.
Likewise, compound systems, 
consisting of a subsystem of 
actual interest and a much larger 
environmental bath, are also 
included as special cases.

Equation (\ref{10}) represents the completely 
general and formally exact solution of 
the dynamics, 
exhibiting the usual
symmetry properties of quantum 
mechanics under time inversion.
Moreover, the right hand side is
a quasi-periodic function of $t$,
giving rise to the well-known quantum 
revival effects \cite{hob71}: 
$\langle A\rangle_{\!\rho(t)}$ 
must return very close to 
$\langle A\rangle_{\!\rho(0)}$ 
for certain, very rare times $t$.

The problem of {\em equilibration}
amounts to the question whether, in 
which sense, and under what conditions 
the expectation value (\ref{10}) 
approaches some constant (time-independent) 
value for large $t$.
Unless this expectation value is constant
right from the beginning, which is {\em not}
the case under generic (non-equilibrium)
circumstances, the above 
mentioned revivals immediately exclude 
equilibration in the strict sense that 
(\ref{10}) converges towards some 
well-defined limit for $t\to\infty$.
On the other hand, ``practical 
equilibration'' in the sense that 
(\ref{10}) becomes virtually indistinguishable 
from a constant value for the overwhelming 
majority of all sufficiently large $t$
has been demonstrated, for instance, in
 \cite{equil,sho11,sho12,rei12,bal16}
under quite weak conditions on 
$H$, $\rho(0)$, and $A$.
In particular, equilibration in 
this sense still admits transient
initial relaxation processes
and is compatible with the above 
mentioned time inversion symmetry
and quantum revival properties.

For the rigorous derivation of these 
results and the detailed requirements
on $H$, $\rho(0)$, and $A$, we
refer to the above mentioned literature.
Here, we confine ourselves to
a complementary, predominantly heuristic 
discussion of the essential points:

Averaging  (\ref{10}) over all 
times $t\geq 0$
yields the result $\langle A\rangle_{\!\rhobar}$,
where the so-called 
diagonal ensemble is defined 
as
\begin{eqnarray}
\rhobar :=\sum_n p_n\, |n\rangle\langle n|
= \sum_n \rho_{nn}(0)\, |n\rangle\langle n|
\ ,
\label{15}
\end{eqnarray}
and where we exploited (\ref{11}) in the last 
step\footnote{If $H$ exhibits degeneracies,
we tacitly choose the eigenvectors
$|n\rangle$ so that $\rho_{mn}(0)$ 
is diagonal within every eigenspace.
Regarding the existence of the time 
average for infinite dimensional 
Hilbert spaces see \cite{rei12}.}.
Given the system equilibrates 
at all (in the above specified sense), 
it follows that (\ref{10}) must 
remain extremely close to 
$\langle A\rangle_{\!\rhobar}$
for the vast majority of all 
sufficiently large times $t$.

Intuitively, the essential mechanism 
is expected to be a ``dephasing''
\cite{equil,sre94,dephasing}
of the oscillating summands on the
right hand side of (\ref{10}):
there must be sufficiently many
different ``frequencies'' $[E_n-E_m]/\hbar$ 
which notably contribute to the sum,
resulting in an approximate cancellation
for most sufficiently large $t$,
provided $H$, $\rho(0)$, and $A$
satisfy certain ``minimal''
conditions:

To begin with, 
some of the oscillating
summands in (\ref{10}) may
assume arbitrary large amplitudes 
by suitably tailoring the
$A_{nm}$'s,
even for otherwise quite harmless
$\rho(0)$ and $H$, thus prohibiting 
equilibration in any meaningful sense.
To exclude such pathologies, a
convenient minimal requirement 
on $A$ turns out to be that it
must represent an experimental 
device with a finite range $\da$ of 
possible measurement outcomes,
where $\da$ is given by 
the difference between the 
largest and smallest 
eigenvalues of $A$.
Furthermore, the resolution limit 
$\dda$ of the considered device 
must be limited to experimentally 
reasonable values compared to its 
working range $\da$.
Quantitatively, all measurements known 
to the present author yield less than 
$20$ significant figures, 
implying that the resolution 
limit $\dda$ must be lower 
bounded by $10^{-20}\da$.
Maybe some day 100 or 1000 significant figures 
will become feasible, but it seems reasonable 
that a theory which does not go very much 
beyond this will do. 
Note that similar restrictions also apply to 
numerical experiments by computer simulations.
We finally remark that the same
or some equivalent assumption on $A$ is, 
at least implicitly, taken for granted 
in all pertinent works in this context,
and it is obvious that considering 
only such observables will be sufficient
for all practical purposes.

Similarly, with respect to $\rho(0)$
it is quite plausible that if two (or more)
level populations $p_n$ in (\ref{11})
with non-degenerate energies $E_n$
are not very small (compared to $\sum_n p_n=1$)
then non-negligible Rabi oscillations may 
arise in (\ref{10}), 
which prohibit equilibration in any 
reasonable  
sense\footnote{This is particularly obvious 
if  $\rho(0)$ is a pure state and hence
$|\rho_{mn}(0)|^2=\rho_{mm}(0)\,\rho_{nn}(0)$.},
even for otherwise 
quite harmless $A$ and $H$.
In other words, 
all level populations must satisfy the 
condition $p_n\ll 1$ apart 
from possibly one exception.
More generally, if $H$ exhibits degenerate eigenvalues
$E_n$, then analogous conditions must be
fulfilled by the populations of the
energy eigenspaces in order to rule out any
non-negligible ``coherent oscillations''
on the right hand side of (\ref{10}).
For similar reasons, not too many of the
``energy gaps'' $E_n-E_m$ in (\ref{10})
may coincide, or if they coincide,
they must contribute with sufficiently 
small weights.
In view of the usually very dense 
and irregular energy spectra, 
the above (or some equivalent)
requirements are commonly taken 
for granted under all 
experimentally relevant conditions.

Given 
$H$, $\rho(0)$, and $A$
satisfy the above ``minimal requirements'', 
there are no further obvious reasons which 
may prevent equilibration via a ``dephasing'' 
of the summands on the right hand 
side of (\ref{10}).
One thus expects that, 
after initial transients have died out, 
the system behaves practically 
indistinguishable from the steady 
state (\ref{15});
deviations are either unresolvably
small (below the resolution 
limit $\dda$) or unimaginably 
rare in time.
All this has been rigorously 
confirmed, e.g., in  
\cite{equil,sho11,sho12,rei12,bal16}.

As an aside we note that the preparation
of an initial condition $\rho(0)$ with a distinct
non-equilibrium expectation value of $A$
at time $t=0$
must actually amount to a quite special selection 
of the terms $\rho_{mn}(0)A_{nm}$ 
(in particular of their complex phases)
on the right hand side of  (\ref{10}) \cite{sre94}.
This issue is in fact also quite closely related to 
a variety of so-called typicality concepts and
results, see  
\cite{llo88,rei15,ham18}.

In the rest of the paper we always tacitly
focus on systems, for which 
the above ``minimal conditions''  
are fulfilled, and hence 
equilibration can be taken for granted.
For the sake of simplicity, we will 
further restrict ourselves to the generic 
case that the energy differences $E_m-E_n$
are non-zero and mutually different for 
all pairs $m\not=n$,
and that 
\begin{eqnarray}
p_n\ll 1
\label{16}
\end{eqnarray}
is fulfilled for all level populations 
in (\ref{11}), i.e., we neglect the
above mentioned generalization that 
there may be one exceptional index 
$n$ which violates (\ref{16}).
Similarly, also our above restriction 
on the energy differences $E_m-E_n$
could in principle still be lifted 
to some degree,
as shown in \cite{sho12,rei12}.

The natural next question is whether
the system exhibits {\em thermalization}, 
that is, whether the long-time average
$\langle A\rangle_{\!\rhobar}$
(see above  (\ref{15}))
is well approximated 
by the pertinent microcanonical 
expectation value, as predicted by 
equilibrium statistical mechanics.
Throughout the present paper, this issue of
{\em whether the system thermalizes 
or not will be largely irrelevant}.
In particular, so-called integrable 
systems and systems exhibiting many 
body localization (MBL), which are 
commonly expected to exhibit equilibration 
but not thermalization
\cite{mor18,gog16,ale16,nan15,gol17}, 
are still admitted.

\section{Typical temporal relaxation}
\label{s2}
Taking for granted equilibration as specified 
above, the main focus of this section is on
the detailed temporal relaxation of the
expectation value (\ref{10}) from its initial
value at time $t=0$ towards the (apparent)
long-time limit $\langle A\rangle_{\!\rhobar}$
(see above  (\ref{15})).

Similarly as in section \ref{s1}, while a 
mathematically rigorous derivation of the 
subsequent results is provided in 
 \cite{rei16,bal17},
the following line of reasoning
amounts to a much shorter, less 
rigorous, but physically more 
instructive alternative derivation.

To begin with, we assume that only some 
large but finite number $D$ of the energy 
levels $E_n$ exhibit non-negligible 
populations $p_n=\rho_{nn}(0)$
(see (\ref{11})) and, without loss of generality, we 
label them so that $n\in\{1,...,D\}$ 
for all those $E_n$.
Accordingly, all other $\rho_{nn}(0)$'s
are approximated as being strictly zero.
For a more detailed, quantitative 
justification of this approximation 
we refer to Appendix A.
The Cauchy-Schwarz inequality 
$|\rho_{mn}|^2\leq \rho_{mm}\rho_{nn}$
then implies that only $m,\,n\leq D$
actually matter in  (\ref{5}), (\ref{10}), (\ref{15}),
i.e.,
\begin{eqnarray}
H & = & \sum_{n=1}^D E_n\, |n\rangle \langle n|
\ ,
\label{14}
\\
\langle A\rangle_{\!\rho(t)}
& = & \sum_{m,n=1}^{\DD}   
 e^{i[E_n-E_m]t/\hh} 
\,
\rho_{mn}(0) A_{nm}
\ ,
\label{17}
\\
\rhobar  & = & 
\sum_{n=1}^D p_n\, |n\rangle\langle n|
=
\sum_{n=1}^D \rho_{nn}(0)\, |n\rangle\langle n|
\ .
\label{18}
\end{eqnarray}
Note that if the number $D$ of non-negligible
level populations were {\em not} large, then
equilibration as discussed in section \ref{s1}
may not be expected in the first place.
On the other hand, (\ref{17}) can be 
shown to approximate (\ref{10}) very 
well under quite general conditions
(see also Appendix A).

The examples of foremost interest
are isolated many-body systems with a
macroscopically well defined energy,
i.e., all relevant energies 
$E_1$,...,$E_{\DD}$ are confined to some 
microcanonical energy window $[E-\Delta E,E]$ 
of microscopically large but macroscopically 
small width $\Delta E$.
Henceforth it is taken for granted that the
considered system is of this type.

The summands with $m=n$ in (\ref{17}) can be readily
rewritten by means of the diagonal ensemble 
from (\ref{18}) as $\langle A\rangle_{\!\rhobar }$,
yielding
\begin{eqnarray}
\langle A\rangle_{\!\rho(t)}
& = & 
\langle A\rangle_{\!\rhobar }
+ {\sum}'
e_{mn}\, a_{mn}
\label{20}
\ ,
\\
e_{mn} & := & e^{i[E_n-E_m]t/\hh}
\ ,
\label{30}
\\
a_{mn}& := & \rho_{mn}(0) A_{nm}
\ ,
\label{40}
\end{eqnarray}
where 
the symbol $\sum'$
indicates a sum
over all $m,n\in\{1,...,D\}$ with $m\not =n$.
Since $D$ is large, the number $D(D-1)$
of those summands is even much larger.

For any given $t$, those very numerous $e_{mn}$'s 
are distributed on the complex unit circle 
according to (\ref{30}).
All of them start out from $e_{mn}=1$ for
$t=0$, and subsequently spread out along 
the unit circle as $t$ increases.
Hence, their distribution on the unit circle
will be highly non-uniform (strongly peaked 
around unity) for small $t$, while they 
are expected to become roughly speaking
uniformly distributed as $t\to\infty$.
More precisely, since the number of 
$e_{mn}$'s is large but finite, their collective 
motion on the unit circle must be quasi-periodic,
i.e., occasional ``recurrences'' 
and other appreciable deviations from a 
uniform distribution necessarily must occur 
for certain, arbitrary large times $t$, 
but they are expected to be
extremely rare and thus safely 
negligible for all practical purposes.

Turning to (\ref{40}), one readily concludes
from the Cauchy-Schwarz inequality that
$|A_{nm}|\leq \norm{A}$, 
where $\norm{A}$ indicates the operator 
norm of $A$ (largest eigenvalue in modulus).
Likewise, one sees that
$|\rho_{mn}(0)|\leq \norm{\rho(0)}\leq 1$,
i.e., all the $a_{mn}$'s 
are distributed inside a circle of
radius $\norm{A}$ in the complex plane.

Note that the matrix elements 
$A_{nm}=\langle n|A|m\rangle$ in (\ref{40}) 
are independent of the energy eigenvalues $E_n$,
while the $e_{mn}$'s in (\ref{30}) are
independent of the corresponding 
energy eigenvectors $|n \rangle$.
Furthermore, only indices $m$ and $n$ 
with macroscopically
small differences $E_n-E_m$
(see below (\ref{18})) and with
$m\not =n$ actually matter in (\ref{20}).
In the absence of any {\em a priori}
reasons to the contrary, one thus expects 
that the quantitative values of the matrix
elements $A_{nm}$ will not be ``correlated''
in any specific way with the $e_{mn}$'s,
see also \cite{ale16,sre99,eth}.
Put differently, how should the observable $A$
``feel'' for example whether or 
not a given pair of 
eigenvectors $|n\rangle$ and $|m\rangle$ 
belongs to a small energy differences $E_n-E_m$ 
in (\ref{30}) without any {\em a priori} 
knowledge about the Hamiltonian $H$ in (\ref{14})\,?
After all, without such extra knowledge, 
the $|n\rangle$'s are orthogonal to each 
other but for the rest may be 
arranged in any way within the high
dimensional Hilbert space under 
consideration.

Similar considerations as for the $A_{nm}$
apply to the matrix elements 
$\rho_{mn}(0)$ in (\ref{40}).

All these arguments suggest that 
both the $e_{mn}$'s and the $a_{mn}$'s 
may be roughly speaking viewed as two 
large sets of pseudorandom numbers,
which are essentially independent of 
each other, implying the approximation
\begin{eqnarray}
\frac{\sum'  e_{mn}\, a_{mn}}{D(D-1)}
=
\frac{\sum'
e_{mn}}{D(D-1)}
\
\frac{\sum'  a_{mn}}{D(D-1)}
\ .
\label{50}
\end{eqnarray}
Indeed, 
since $D(D-1)$ is the number of 
summands in each of the three 
sums in (\ref{50}),
the left hand side amounts to
the correlation of the $e_{mn}$'s and
the $a_{mn}$'s, which, for statistically
independent random numbers and $D\to\infty$, 
is known to converge (with probability $\to 1$)
towards the product of the two 
mean values on the right hand side.
Qualitatively, somewhat similar ideas have
also been developed in  \cite{dephasing},
but the quantitative details were quite
different.

Concerning the above 
justification of (\ref{50}),
our first side remark is that the $e_{mn}$'s and 
the $a_{mn}$'s are actually only required to be
uncorrelated, which is strictly speaking
a weaker condition than being independent.
Second, we note that the $e_{mn}$'s need not be
uniformly distributed on the unit 
circle\footnote{Also 
in probability theory, two random
variables may well be statistically 
independent (or uncorrelated),
no matter how each of the two 
single variables 
is distributed. One (or both) of them
may even be non-random
(corresponding to a delta-distribution),
in which case the independence 
property is always trivially fulfilled.}.
Third, focusing on the $a_{mn}$'s alone,
it is not necessary that they are uncorrelated 
or independent from each other,
and likewise for the $e_{mn}$'s.

This heuristic approximation in (\ref{50})
will be the key 
ingredient of our subsequent line of reasoning.
Further arguments in support of it are:
(i) It amounts to an exact identity for $t=0$.
(ii) Likewise, upon averaging over all times 
$t\geq 0$ and taking for granted 
that all energies $E_n$ are non-degenerate
(see above (\ref{16})), 
one can show that (\ref{50})
becomes an exact identity.

The first sum on the right hand side of (\ref{50})
can be rewritten by means of (\ref{30}) 
as
\begin{eqnarray}
{\sum}' e_{mn}
& = & 
\sum_{m,n=1}^D 
e^{i[E_n-E_m]t/\hh} - \sum_{n=1}^D e^0
\nonumber
\\
& = &
| D \phi(t)|^2 - D
\ ,
\label{60}
\\
\phi(t)
& := &
\frac{1}{\DD}\sum_{n=1}^{\DD}e^{i E_n t/\hh} 
=\frac{1}{\DD}\mbox{Tr}\left\{e^{iHt/\hh}\right\}
\ .
\label{70}
\end{eqnarray}
Likewise, the last sum in (\ref{50})
can be rewritten by means of (\ref{40}) as
\begin{eqnarray}
{\sum}'  a_{mn}
=
\sum_{m,n=1}^D  \rho_{mn}(0) A_{nm} 
- \sum_{n=1}^D \rho_{nn}(0) A_{nn}
\label{75}
\end{eqnarray}
and with (\ref{17}), (\ref{18})
it follows that
\begin{eqnarray}
{\sum}'  a_{mn}
& = &
\langle A\rangle_{\!\rho(0)}
- \langle A\rangle_{\!\rhobar }
\ .
\label{80}
\end{eqnarray}

Upon introducing (\ref{50})-(\ref{80})
into (\ref{20}),
we finally obtain as our first 
main result the approximation
\begin{eqnarray}
\langle A\rangle_{\!\rho(t)} & = & 
\langle A\rangle_{\!\rhobar }
+
F(t)\, \left[\langle A\rangle_{\!\rho(0)} 
- \langle A\rangle_{\!\rhobar }\right]
\ ,
\label{90}
\end{eqnarray} 
where $F(t):=(D|\phi(t)|^2-1)/(D-1)$.
Since $D\gg 1$ this yields the very accurate 
approximation
\begin{eqnarray}
F(t)=|\phi(t)|^2 \ ,
\label{100}
\end{eqnarray}
where $\phi(t)$ is given by (\ref{70})
and therefore may be interpreted as 
the Fourier transform of the 
spectral density of $H$.

The key ingredient for the derivation 
of (\ref{90}) was the heuristic 
approximation (\ref{50}).
While it makes the derivation short and
physically instructive, a more rigorous 
justification of (\ref{50})
seems very difficult.
On the other hand, the very same formula
(\ref{90}) can also be rigorously obtained
by means of a technically very different,
more arduous and less instructive 
approach, see  \cite{rei16,bal17}, using 
averages over unitary transformations,
under which the locality properties of a given
Hamiltonian are in general not preserved
(see also sections \ref{s3} and \ref{s4}).

Upon comparison with (\ref{70}) we see that
$F(t)$ in (\ref{100}) quantifies the above
discussed distribution of the $e_{mn}$'s
on the complex unit circle.
In particular, one readily 
finds that $F(0)=1$ and $0\leq  F(t)\leq 1$ 
for all $t$.
Moreover, the following properties 
of $F(t)$ 
were derived previously in  \cite{rei16}:
(i) $F(t)$ remains negligibly small
for the vast majority of all 
sufficiently large times $t$, provided 
the maximal degeneracy of the energies
$E_1,...,E_D$ is much smaller 
than $D$ (see also above  (\ref{16})).
The extremely rare exceptional $t$'s 
are inherited from the above mentioned 
quasi-periodic motion of the $e_{mn}$'s
on the unit circle.
Our main result (\ref{90}) thus
captures at least qualitatively correctly 
the decay from  the initial 
expectation value
$\langle A\rangle_{\!\rho(0)}$
towards the long-time average
$\langle A\rangle_{\!\rhobar }$,
and also the well-known quantum 
revivals at arbitrarily large but 
exceedingly rare times \cite{hob71}.
(ii) Denoting by $\Omega(E)$ the number of
energies $E_n$ below $E$, by $\kB$ and
$S(E):=\kB\ln \Omega(E)$ 
Boltzmann's constant and entropy,
respectively, and by $T:=1/S'(E)$
the corresponding formal temperature,
one can often approximate the sum 
in  (\ref{50}) by an integral over
a suitably smoothened level density,
yielding the approximation
\begin{eqnarray}
F(t)=1/[1+(t\,\kB T/\hbar)^2] \ .
\label{110}
\end{eqnarray}
As may have been expected, the above 
mentioned quasi-periodicities of $F(t)$
and the concomitant quantum revivals
get lost within such a continuum 
approximation.
We also note
that $T$ and $S(E)$ can be identified 
with the usual temperature and entropy
of the thermalized system (at energy $E$), 
provided the system does approach
thermal equilibrium for large times
(see end of section \ref{s1}).

In the opposite case of a 
non-thermal long-time limit, 
$T$ and $S(E)$ are usually 
still well defined formal 
quantities, but without an 
immediate physical meaning.
Rather, they may be viewed as 
the equilibrium temperature and 
entropy of some auxiliary initial 
state $\rhoaux (0)$, which
does exhibit thermalization,
and whose energy expectation 
value $\mbox{Tr}\{\rhoaux (0) H\}$
is identical to the ``true'' 
system energy 
$E:=\mbox{Tr}\{\rho(0) H\}$.
In particular, such a $\rhoaux (0)$
always exists (for instance the
microcanonical ensemble), and
hence (\ref{110}) remains valid
even for non-thermalizing initial 
states $\rho(0)$.
The only prerequisite is that
the thermal equilibrium properties 
of $H$ are
``as usual'', i.e., the density of states
is very high and grows very fast with $E$.

A further implication
of (\ref{70}) and (\ref{100}) is 
that $F(-t)=F(t)$ for all $t$.
Hence, the fundamental symmetry properties
of quantum mechanics under time inversion
mentioned below (\ref{10}) are still 
maintained by (\ref{90}).
Remarkably, the time 
inversion symmetry of (\ref{90})  
even persists in cases where
it is broken in the microscopic 
quantum dynamics, e.g., due to an 
external magnetic field.
This is reminiscent of the second law of 
thermodynamics, which also remains valid
for systems with a magnetic field
and thus with broken microscopic 
time inversion symmetry.

\section{Exceptional cases}
\label{s3}
In this section, we collect the main 
{\em a priori} reasons
announced above  (\ref{50}), which may
invalidate the approximation (\ref{50})
and hence our main result (\ref{90}).

To begin with, we note that 
$\langle n|[H,A]|m\rangle = (E_n-E_m)A_{nm}$,
where $[H,A]$ is the commutator 
between the Hamiltonian
(\ref{14}) and the observable $A$.
If $A$ is a conserved quantity
it satisfies $[H,A]=0$, implying that
$A_{nm}=0$ whenever $E_n\not=E_m$.
If we now slightly perturb the Hamiltonian
under consideration, one can 
infer from ordinary perturbation theory
(for extremely small perturbations)
or more sophisticated non-perturbative 
methods \cite{fyo96} 
(for moderately small perturbations)
that the new matrix elements $A_{nm}$ 
in the basis of the perturbed Hamiltonian
are non-negligible only for
relatively small $E_n-E_m$.
With reference to the new, slightly
perturbed system, the observable
$A$ may thus be called ``almost conserved'',
still exhibiting a significant
correlation between the energy
differences $E_n-E_m$ and the 
magnitude of the matrix elements
$A_{nm}$.
Hence, also the
$e_{mn}$'s in (\ref{30}) and the
$a_{mn}$'s in (\ref{40}) will
be correlated and the argument 
below (\ref{50}) breaks down.
One thus expects that the 
temporal relaxation of such an almost 
conserved quantity will be 
slower than predicted by (\ref{90}).

Important examples are the energies 
of two weakly coupled subsystems
(of an isolated compound system),
or the total momentum 
of an isolated system, 
such as a simple gas in a box,
which is not conserved 
due to momentum exchange 
with the system boundaries
(and similarly for the total 
angular momentum).
All these observables then amount 
to almost conserved quantities since
they represent ``volume'' properties
(extensive quantities),
which only can change through 
``surface'' effects (exchange of 
energy, momentum etc. via
``particle-wall interactions'').
Our present theory only applies if
such quantities assume their
equilibrium value right from the
beginning (e.g., the total momentum must be zero), 
or if they can be approximated as being strictly 
conserved (e.g., the weak coupling between 
subsystems is ``switched off'').
Put differently, this is a first instance
where we see that macroscopic 
transport in the sense of 
section \ref{s0} must be excluded.

An analogous breakdown of (\ref{50})
and hence of (\ref{90}) is expected if
$\rho(0)$ is an ``almost conserved''
quantity. 

Next, let us replace the original 
$H$ from (\ref{14}) by the transformed 
Hamiltonian
\begin{eqnarray}
H_U:=U\, H\, U^\dagger \ ,
\label{120}
\end{eqnarray}
where $U$ is an arbitrary but
fixed unitary transformation.
In other words, the eigenvalues of $H_U$ 
are still given by $E_n$, while the
eigenvectors are now $U|n\rangle$
instead of $|n\rangle$.
Accordingly, the original definition 
$\rho_{mn}(0):=\langle m|\rho(0)|n\rangle$
in (\ref{17})-(\ref{80}) must be replaced by 
$\rho_{mn}(0):=\langle m|U^\dagger \rho(0) U|n\rangle$,
and analogously for the definitions of
$A_{nm}$ and of $\rhobar$ in (\ref{18}).
In the final result (\ref{90}), the initial 
value $\langle A\rangle_{\!\rho(0)}$ 
as well as the function $F(t)$ are 
not affected by such a unitary transformation, 
while the quantitative value of the
long-time average $\langle A\rangle_{\!\rhobar}$
may in general change.
Similarly, the $e_{mn}$'s in (\ref{30}) 
are independent of $U$, while
the $a_{mn}$'s in (\ref{40}) are
typically ``redistributed'' in a 
very complicated way.
Therefore,  (\ref{50}) is expected
to be satisfied in very good 
approximation for most $U$'s.
A more detailed verification of this
expectation is provided in 
 \cite{rei16,bal17}.
The key point is that this finding is
independent of whether (\ref{50})
was satisfied by the original 
Hamiltonian $H$ in (\ref{14})
or not.

In conclusion, (\ref{90}) cannot be
correct if the temporal relaxation,
encapsulated by the $U$ independent
function $F(t)$, is notably different
for the ``true'' Hamiltonian $H$ than 
for most other Hamiltonians $H_U$.

One readily sees that the latter criterion,
in particular, also excludes the previously discussed cases 
when $A$ or $\rho(0)$ is an almost 
conserved quantity.

\section{Restriction to transportless relaxation}
\label{s4}
A pivotal feature of almost all physical systems
of interest is that they can be very well described in
terms of some 
``elementary constituents'' (atoms, molecules, 
quasiparticles etc.),
which are reasonably localized in space and
whose interaction is of short range.
Formally, the model Hamiltonian $H$ is 
thus composed solely of so-called 
{\em local operators}.
Only in such cases it makes sense to ask for the 
amount of energy, charge, particles etc. 
within some subdomain of the system:
If the considered volume is not too small
then the interaction with the rest of the system 
is weak and can be approximately ignored
(surface effects are small compared to 
volume contributions).
In other words, local densities are 
reasonably well-defined concepts.
Since they are usually ``local descendants''
of some globally conserved quantities (energy, 
charge, particle numbers etc.) their content 
within a given volume can only change via 
{\em transport currents} through the boundaries 
of that volume.

As discussed in section \ref{s1}, all those local
densities will equilibrate towards certain 
(approximately) steady values after 
sufficiently long times.
If all local densities for a given initial state
$\rho(0)$ agree (at every point in space 
and in sufficiently good approximation)
with the corresponding equilibrium values, 
then $\rho(0)$ is called a {\em macroscopically 
homogeneous initial state}.
The word ``homogeneous'' refers to the fact 
that the densities after equilibration
are indeed spatially homogeneous in many 
examples of interest.
For simplicity, we tacitly focus on
such situations in the following 
discussion.
However, {\em analogous conclusions 
remain valid even when the equilibrated 
densities are actually inhomogeneous}.
(It is only the naming which becomes 
``wrong'', not the argument).
The word ``macroscopic'' refers to the 
fact that the very concept of a density 
or a transport current breaks down 
on microscopic length scales.
(For instance, the number of atoms
within a small volume should be 
well approximated by the corresponding 
particle density times the volume.
If the volume is so small that it 
only contains a few atoms, this is 
no longer true. Put differently,
the microscopically discrete particles
are no longer well described by
a continuum approximation in terms of 
densities and concomitant currents.)

In real systems, the equilibration of initial 
inhomogeneities via the above mentioned 
transport currents takes an increasingly 
long time over increasingly large distances.
On the other hand, the function $F(t)$
from (\ref{70}) and (\ref{100}),
which governs the temporal relaxation 
in (\ref{90}), is independent of the
initial state and thus of the distance 
between possible inhomogeneities.
Moreover, the characteristic
time scale, predicted, e.g., by 
(\ref{110}) is very short 
($\hbar/\kB T\simeq 26\,$fs at room temperature).
In other words, (\ref{90}) must be invalid
for initial conditions which give 
rise to significant spatial inhomogeneities 
on macroscopic scales. 

The underlying {\em a priori} reason
(see section \ref{s3}) is as follows.
In contrast to $H$ (see above), most 
transformed Hamiltonians $H_U$ in (\ref{120})
can no longer be interpreted 
as a description of certain basic constituents 
(atoms etc.) which are spatially well 
localized and exhibit short range interactions,
nor can they any longer be rewritten as (sums of)
{\em local operators}.
Hence, local densities and transport currents
are not any more well defined, and the
very same initial conditions $\rho(0)$, 
which entailed spatial inhomogeneities when 
dealing with $H$, are no longer expected to 
equilibrate particularly slowly when $H_U$ 
governs the dynamics.
Hence the ``exclusion criterion'' at the
end of section \ref{s3} applies
to such a system Hamiltonian $H$.

It is interesting to consider the same 
thing from yet another viewpoint.
Namely, one readily sees from the discussion 
below (\ref{120}) that instead of replacing 
$H$ by $H_U$ (while leaving $\rho(0)$ and
$A$ unchanged), one could as well
keep $H$ unchanged and replace $\rho(0)$
and $A$ by 
$\rho_U(0):=U^\dagger \rho(0) U$
and 
$A_U:=U^\dagger A U$,
respectively.
In other words, only the initial state 
and the specific observable under 
consideration are changed, whereas 
local densities etc. are represented 
by the same operators
before and after the transformation,
and, in particular, still remain 
perfectly well defined concepts even 
in the transformed setup.
For any given such invariant operator $B$,
one can show along the lines of 
 \cite{rei15} that the 
initial expectation value 
$\langle B\rangle_{\!\rho_U(0)}$ 
is practically indistinguishable from
the pertinent equilibrium value
$\langle B\rangle_{\!\rhobar}$
for most  $U$'s.
In particular, $B$ may quantify the 
amount of energy (or charge etc.) 
within a macroscopically small but 
microscopically still not too small 
volume $V$, and thus $B/V$ 
accounts for the
corresponding density at the
location of that volume.
The same remains true simultaneously 
for several different observables 
$B_1,..,B_K$, where $K$ 
may be sufficiently large to
specify the entire spatial dependence 
of the densities within any experimentally 
resolvable resolution.
As a consequence, most $\rho_U(0)$'s 
must be (approximately) homogeneous
and hence their relaxation
(under $H$) is not expected to be 
particularly slow.

In conclusion, systems with short range interactions
in combination with initial conditions,
which give rise to non-negligible spatial
inhomogeneities on macroscopic scales,
must be excluded in (\ref{90}).
Put differently,  the total energy, (angular) 
momentum, particle numbers etc. within 
any macroscopic part of the system must 
remain constant during the entire relaxation 
process.
Accordingly, the relaxation process must 
not entail any significant transport currents, 
caused by some unbalanced local densities.

For instance, such a transportless relaxation
scenario often arises quite naturally when 
the system Hamiltonian and the initial 
non-equilibrium state do not exhibit any 
spatial inhomogeneities on macroscopic 
scales.
Strictly speaking, one also
has to exclude the possibility of
spontaneous symmetry breaking 
during relaxation,
initial states with non-vanishing 
total momentum (resulting in transport 
through system boundaries),
etc., see also section \ref{s3}.

In case of notable spatial inhomogeneities,
it may still be possible to approximately 
partition the system into sufficiently small,
non-interacting subsystems and then
describe the relaxation within each 
of them by (\ref{90}).
Essentially, this is tantamount to the
well established concept of local 
equilibration.
Usually, this local equilibration is much
faster than the subsequent, global equilibration
of the small subsystems relatively to each other.
The latter, slow processes are no longer 
covered by our theory (\ref{90}).
In turn, the clear-cut separation
of the two time scales usually admits
some Markovian approximation for the slow
processes, resulting in an exponential
decay, whose timescale still depends
on many details of the system.
For similar reasons, also correlation
and entanglement properties
of spatially well separated regions 
are beyond the realm of our present theory;
very roughly speaking, they may be viewed 
as being governed by {\em transport of information}, 
whose propagation speed is limited, e.g., 
by Lieb-Robinson bounds \cite{gog16,bra06}.

Closely related further generalizations of 
the above local equilibration paradigm
are the concepts of hindered equilibrium,
quasi-equilibrium, metastability,
and, above all, prethermalization
\cite{mor18,ber04,moe08,gri12,lan16,mal18}.
The first three concepts play a crucial
role for instance in chemical reactions with
long-lived intermediates,
or in quantum systems exhibiting ``glassy behavior'' 
\cite{mor17,tur18},
while the concept of prethermalization refers,
e.g., to a fast but only partial 
thermalization of a certain 
subset of modes, 
(quasi-)particles\footnote{In general,
quasiparticles are expected to become 
a meaningful concept
only after prethermalization \cite{moe08}.},
or other generalized degrees of 
freedom \cite{rei16}.

More formally, the latter cases have their origin in 
certain almost conserved quantities of the pertinent
Hamiltonian $H$, which significantly slow down
some intermediate steps of the temporal relaxation,
while the same is no longer true for most of the 
transformed Hamiltonians $H_U$ within the framework 
discussed at the end of section \ref{s3}.

As already mentioned, analogous conclusions
remain valid even when the equilibrated 
densities are actually inhomogeneous,
provided all of them are (approximately) 
equal to the initial densities.
The only indispensable prerequisite is
the absence of transport during relaxation.
This case is of particular interest
when {\em the system is composed of
a small subsystem of actual interest 
and a bath}.
Usually the bath can be considered as 
equilibrated right from the beginning, 
hence the decisive question is
whether all densities in the small 
subsystem remain (practically) 
unchanged during the equilibration 
process.
In particular, if the subsystem
is so small that no meaningful
local densities can be defined,
then the above considerations
no longer imply that some initial 
conditions must be excluded
{\em a priori}.
In turn, if the subsystem is not 
small and all transport currents 
are still excluded, one expects a largely 
similar relaxation behavior in the
presence and in the absence 
of the bath.

\section{Comparison with experiments}
\label{s6}
As recognized in the preceding section \ref{s4}, 
an indispensable prerequisite of our present 
theory
is that the initial non-equilibrium 
state must be spatially homogeneous.
Though most published experiments 
on equilibration and thermalization 
admittedly do not fulfill this 
requirement, there still exists a 
considerable number which do fulfill it.

A variety of such experimental (as well
as numerical) data from the literature
have been demonstrated already in  
\cite{rei16,bal17} to agree remarkably
well with the theoretical 
predictions in (\ref{90}) and (\ref{110}).
It is worth mentioning that most of those
data have not been quantitatively explained 
by any other analytical theory so far.

Note that the relevant time 
scale $\hbar/\kB T$ in (\ref{110})
is approximately $26\,$fs at room 
temperature.
In many cases, such extremely fast
processes may be experimentally 
difficult to observe, or they have 
simply not been looked for until now.
In particular, spatially inhomogeneous
initial conditions usually exhibit
a much slower relaxation, but they are 
not covered by our present theory.
On the other hand, for systems at extremely low 
temperatures, 
such as
atomic Bose gases, 
the relevant time scale $\hbar/\kB T$ will be more easily
accessible, hence these are
promising candidates
for a comparison with our present 
theory \cite{rei16,bal17}.
Finally, the relaxation dynamics 
near a quantum critical point is 
known to be governed by 
the very same time scale $\hbar/\kB T$
under very general conditions,
i.e., independently of any further 
microscopic details of the system
\cite{sac11}.

For a concrete experimental (or numerical)
setup at hand, the value of
$\langle A\rangle_{\!\rho(0)}$ 
in (\ref{90}) is sometimes quite obvious,
but more often its quantitative determination
is very difficult by purely theoretical 
means, and likewise for
the long-time average 
$\langle A\rangle_{\!\rhobar}$
in (\ref{90}).
On the one hand, to analytically determine
those values is not a main issue of our 
present work.
On the other hand, even the experimental 
data themselves are often reported 
in arbitrary units.
Therefore, the quantitative values
of $\langle A\rangle_{\!\rho(0)}$ 
and $\langle A\rangle_{\!\rhobar}$
in (\ref{90}) usually must be taken over 
from the experiment (or the numerics),
hence the only remaining
parameter of the theory is the 
temperature $T$ in (\ref{110}).
Once again, the relevant temperature value, 
as discussed below (\ref{110}), is often 
not available as an experimentally determined 
quantity,  and hence must be estimated indirectly 
or treated as yet another 
fit parameter \cite{rei16,bal17}.

\begin{figure}
\epsfxsize=0.95\columnwidth
\epsfbox{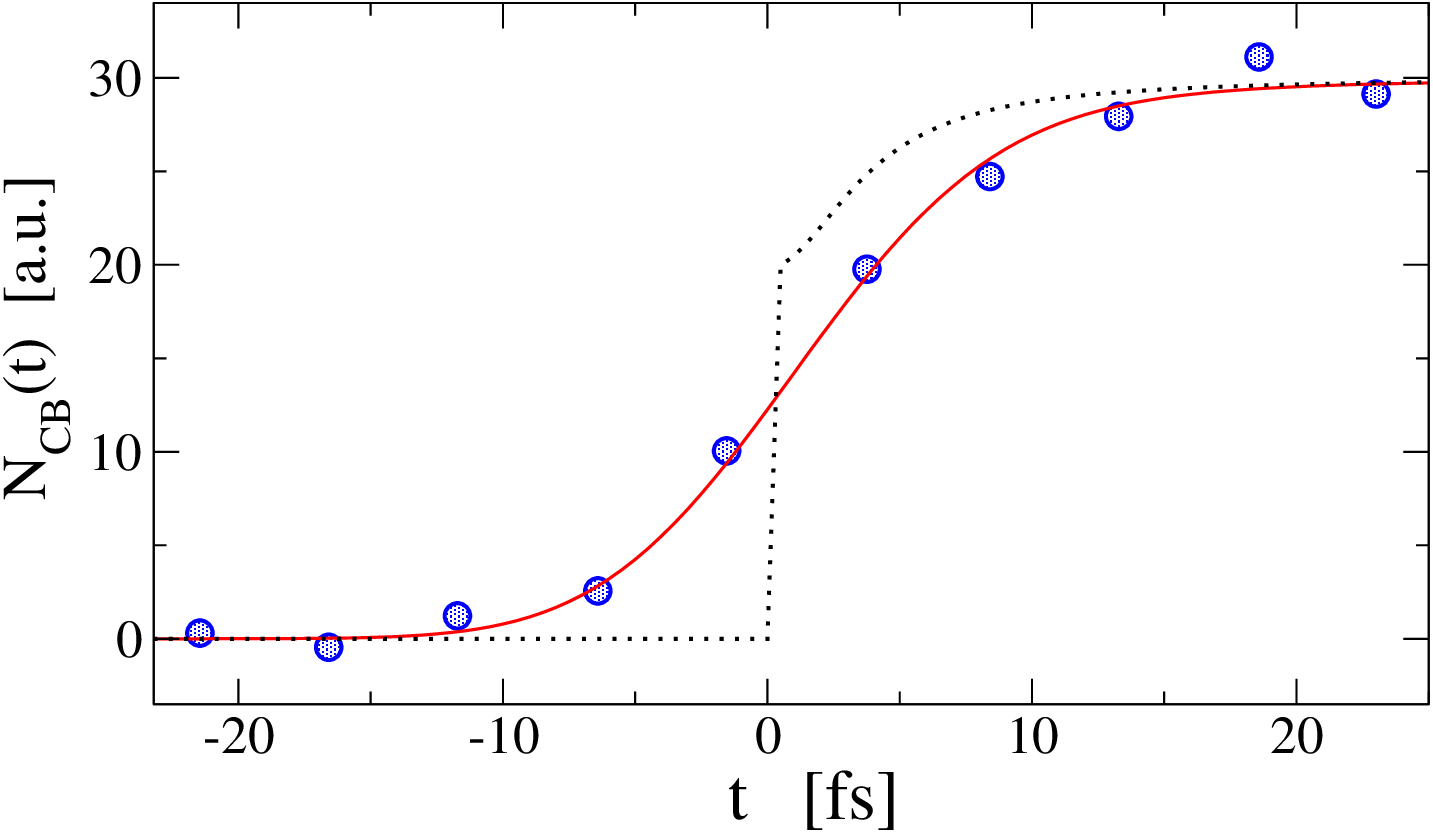}
\caption{\label{fig1}
Symbols: Experimental pump-probe
data from figure 3 in \cite{gie15},
representing the number of electrons $N_{CB}(t)$ in the conduction 
band of a graphene monolayer 
(in arbitrary units).
Dotted: Theoretical prediction 
(\ref{90}), (\ref{110}) for the
observable $\langle A\rangle_{\!\rho(t)}=N_{CB}(t)$
with $T=2000$\,K, 
complemented by 
$\langle A\rangle_{\!\rho(t)}=0$
for $t<0$,
$\langle A\rangle_{\!\rho(0)}=20$,
and
$\langle A\rangle_{\!\rhobar}=30$
(see  (\ref{6.1}) and main text).
Solid: Convolution of the dotted line 
with a Gaussian 
of standard deviation $5.5$\,fs, 
accounting for the finite widths of the 
pump and probe laser pulses.
}
\end{figure}

In the remainder of this section, we 
focus on one of the rare examples,
for which the pertinent temperature 
in (\ref{110}) is experimentally available.
Namely, we consider the pump-probe 
experiment from  \cite{gie15}, where 
the electron gas in a graphene monolayer is 
excited by an ultrashort ``pump'' laser 
pulse, and then its re-thermalization is monitored 
by a second ``probe'' pulse, yielding
the number of electrons in the 
conduction band $N_{CB}$,
see also figure \ref{fig1}.
In other words, the observable $A$
in (\ref{90}) is chosen so that
$\langle A\rangle_{\!\rho(t)}=N_{CB}(t)$.
A more detailed modeling of the actual
observable $A$ corresponding to the
experimental measurement procedure
would be quite difficult, but fortunately 
is not needed\,!

Prior to the pump pulse,
the system is at room temperature 
and $\langle A\rangle_{\!\rho(t)}=N_{CB}(t)$
is known to be negligibly small \cite{gie15};
i.e., $N_{CB}(t)=0$ for $t<0$.
At time $t=0$, the pump pulse
suddenly excites a certain number
$\langle A\rangle_{\!\rho(0)}=N_{CB}(0)$
of electrons into the conduction band
(hence the discontinuity of the dotted 
line in figure \ref{fig1}).
Subsequently, these excited
electrons generate secondary 
electron-hole pairs via impact 
ionization (inverse Auger 
scattering) so that 
$\langle A\rangle_{\!\rho(t)}=N_{CB}(t)$
further increases \cite{gie15}.
If the electron gas were strictly
isolated from the rest of the world
(as assumed in our theory),
it would approach a new thermal 
equilibrium with some temperature 
$T$.
Identifying the corresponding long-time 
average of $N_{CB}(t)$
with $\langle A\rangle_{\!\rhobar}$
in (\ref{90}), one can deduce from
figure 6a in  \cite{mal17} the
estimate
\begin{eqnarray}
\langle A\rangle_{\!\rhobar}/ \langle A\rangle_{\!\rho(0)} 
\simeq 1.5
\ .
\label{6.1}
\end{eqnarray}
In particular, the corresponding
electron gas temperature in figure 6e 
of \cite{mal17} is comparable
to the experimentally relevant value
(see below).
However, in the actual experiment, 
there is -- besides the dominating electron-electron 
interactions -- also a relatively 
weak interaction via electron-phonon 
scattering with the atomic ``background-lattice''
of the graphene layer, 
resulting in a relatively slow relaxation 
of the electron-lattice compound 
towards a thermal equilibrium state
of the total system, which is different
from the above mentioned hypothetical
equilibrium of the electron gas alone, 
and which is not covered by our 
present theory
(the energy of the electron gas is an
almost conserved quantity, 
see section \ref{s3}).
Experimentally, one observes that
the phonon effects are still approximately
negligible for times up to about
$t=25\,$fs, while the electron gas
already approximately thermalizes.
Therefore, only times up to
$t=25\,$fs have been included 
in  figure \ref{fig1}.
In turn, one can deduce from
Figure 4
in the Supplemental 
Material of \cite{gie15}
that the corresponding electron 
temperature $T$ in (\ref{110}) 
is approximately $2000\,K$.

The resulting theoretical prediction
is indicated as dotted line in 
figure \ref{fig1} and does not agree
very well with the experimental data.
The quite obvious reason is
that while both laser pulses are 
extremely short in the experiment,
their duration is still 
not negligible compared to 
the relaxation time scale
of the electron gas.
Theoretically, we roughly take into 
account the finite widths of both 
pulses by convoluting our above
prediction with a Gaussian 
of standard deviation $5.5$\,fs.
The latter value for the combined 
widths of both pulses has been 
experimentally determined,
as detailed in the Supplemental
Material of \cite{gie15} (see
last paragraph of page 3 therein).
The so obtained solid line in
figure \ref{fig1} agrees very well 
with the experimental findings,
especially in view of the fact
that, apart from the unknown 
units of the experimental data, 
there remains no free fit parameter 
in the underlying theory.

With respect to the probe pulse,
the above convolution with a Gaussian
seems an intuitively quite plausible
modeling of the ``smeared out'' 
time point $t$ of the experimental measurement.
With respect to the pump pulse,
it represents a rather poor ``effective 
description'' since our entire theoretical 
approach becomes strictly speaking 
invalid when the duration of the initial 
perturbation becomes comparable 
to the relaxation time \cite{rei16}.
One the other hand, it still seems reasonable
to expect that the finite widths of the pump
and of the probe pulses will have roughly
comparable effects on the measurement 
outcome.
Alternatively, one may imagine that
the probe pulse is indeed very sharply
peaked in time, but the location of the 
delta-peak is slightly different for spatially
different regions on the graphene 
monolayer, and that those regions
interact only very weakly with 
each other.

\section{Amended theory of transportless relaxation}
\label{s5}
As already mentioned in section \ref{s1},
generic many-body systems
exhibit an extremely dense energy 
spectrum: for a macroscopic system
with $f\gg 1$ degrees of freedom,
the distance between neighboring
energy levels is  
exponentially small in $f$.
Hence, even for an initial state
$\rho(0)$ with a macroscopically well
defined energy, there is still
an exponentially large number of
energy levels $E_n$ which 
{\em a priori} may possibly be populated 
with a non-negligible probability 
$p_n$ in (\ref{11}).
Moreover, it seems reasonable to assume 
that it is impossible to experimentally 
realize initial states $\rho(0)$ with
appreciable populations $p_n$ of only 
a few energy levels.
(The opposite case essentially amounts 
to a Schr\"odinger cat and usually rules
out equilibration in the sense of 
section \ref{s1} right from the beginning).
In view of $\sum_n p_n=1$ it follows that 
every single $p_n$ must be extremely 
small (usually exponentially small 
in $f$), see also (\ref{16}).
All these assumptions are tacitly
taken for granted in textbook statistical 
physics and also in all what follows.

Even when every single level population
$p_n$ is very small, some of them 
may still be even much smaller than 
others
(for instance those with energies $E_n$ 
far outside the microcanonical energy
window $[E-\Delta E, E]$ mentioned 
below (\ref{18})).
An important implicit assumption 
of the approach from section \ref{s2} 
is that some of them are actually
negligible (can be approximated as 
being strictly zero), while all the others
can be treated on an equal footing.
But in practice, the quantitative choice of 
the threshold between negligible and 
non-negligible $p_n$'s is often
somewhat ambiguous. Moreover, all
the remaining non-negligible $p_n$'s are 
usually still far from being approximately 
equally large, hence it is not obvious 
why the larger ones should not
play in some sense a more important 
role than the smaller ones. 
The main objective of this section is to 
amend the approach from section \ref{s2} along 
these lines.
Accordingly, we no longer work 
with  (\ref{14})-(\ref{18}) 
but rather return to the original 
equations (\ref{5})-(\ref{15}).

\subsection{Setting the stage}
\label{s51}
Our starting point is the following
property of the dynamics (\ref{10}), 
which is intuitively quite plausible 
and rigorously derived in Appendix A:
Consider an arbitrary but fixed $\rho(0)$
with level populations $p_n$ as defined in (\ref{11}).
Next we choose a set of ``auxiliary populations'' 
$\tilde p_n$, which satisfy $\tilde p_n\geq 0$
and $\sum_n\tilde p_n=1$, but otherwise may 
still be arbitrary.
Then there exists a corresponding 
``auxiliary density operator'' 
$\tilde \rho(0)$
with level populations 
\begin{eqnarray}
\tilde \rho_{nn}(0)=\tilde p_n
\label{201a}
\end{eqnarray}
and with the property that
\begin{eqnarray}
\langle A\rangle_{\!\rho(t)} = \langle A\rangle_{\!\tilde\rho(t)}
\label{201}
\end{eqnarray}
is satisfied in very good approximation
for arbitrary $t$ and $A$ on condition that
\begin{eqnarray}
\sum_n|p_n-\tilde p_n| \ll 1 \ .
\label{201b}
\end{eqnarray}
Taking for granted (\ref{201b}), 
we thus can and will work with 
$\tilde \rho(t)$ instead of $\rho(t)$ 
in the following.
In particular, sufficiently small
$p_n$'s can now be safely replaced 
by strictly vanishing $\tilde p_n$'s.
Moreover, also the remaining non-negligible
$p_n$'s may be ``redistributed'' among 
the $\tilde p_n$'s within the limits
imposed by (\ref{201b}).
Since every single $p_n$ is usually
still extremely small (see above),
quite significant changes of many
level populations are still 
admissible along these lines.
(However, choosing all
the non-vanishing
$\tilde p_n$'s equally large is usually still
impossible without violating 
(\ref{201b}).)
The explicit form of $\tilde \rho(t)$ 
is provided in Appendix A, showing that
$\tilde \rho(t)$ still closely resembles 
$\rho(t)$ if (\ref{201b}) is fulfilled. 
Moreover, whenever $\rho(t)$ is a
pure state, also $\tilde\rho(t)$ 
will be pure.

Incidentally, the above approximation (or the 
more precise version in (\ref{a10})) seems to
be a quite interesting new result on its own,
that may also be of use for instance
in the context of quantum information.

In a second step we assume that the 
Hamiltonian which governs the time 
evolution of $\tilde\rho(t)$ 
is not any more given by (\ref{5})
but rather by
\begin{eqnarray}
\tilde H:=\sum_n\tilde E_n\,|n\rangle\langle n| 
\ .
\label{204}
\end{eqnarray}
As a result, one again finds that (\ref{201}) 
remains a very good approximation on 
condition that
\begin{eqnarray}
t & \ll &
\tmax:=\frac{\hbar}{\max\limits_{n\in I}|\tilde E_n-E_n|} 
\label{203} \ ,
\end{eqnarray}
where $I$ denotes the set of indices $n$ 
with non-vanishing level populations 
$\tilde p_n$,
\begin{eqnarray}
I & := & \{n \, |\, \tilde p_n>0\,\} \ .
\label{203a}
\end{eqnarray}
Intuitively, this finding
appears quite plausible upon a 
closer look at the time evolution
of $\rho(t)$ in (\ref{10})
and the analogous formula for 
$\tilde\rho(t)$.
A more detailed derivation is 
provided in Appendix B.

\subsection{Main idea and assumptions}
\label{s52}
Very roughly speaking, the key idea
is to tailor suitable degeneracies 
of the modified energies $\tilde E_n$'s 
in (\ref{204}) so that the 
probabilities $\tilde p_n$ are
equally distributed among the 
different eigenspaces.
More precisely, the set $I$ in (\ref{203a})
must be partitioned into $M$ disjoint 
subsets $I_1,...,I_M$ with the property 
that all energies $\tilde E_n$ with 
$n\in I_\mu$ are equal, say
\begin{eqnarray}
\tilde E_n := E'_\mu\ \mbox{for all $n\in I_\mu$} \ ,
\label{241}
\end{eqnarray}
and the concomitant ``eigenspace populations''
\begin{eqnarray}
p'_\mu :=\sum_{n\in I_\mu} \tilde p_n 
\label{240}
\end{eqnarray}
are equal for all $\mu=1,...,M$.
Since $\sum_{n\in I} \tilde p_n =1$ 
we can conclude that
\begin{eqnarray}
\sum_{\mu=1}^M p'_\mu & = & 1 
\label{242}
\end{eqnarray}
and thus
\begin{eqnarray}
p'_\mu & = & 1/M 
\label{243}
\end{eqnarray}
for all $\mu=1,...,M$.

In the above described construction, 
two further constraints have
to be taken into account for 
reasons that will become clear shortly:
(a) The number of subsets $M$ must
be large,
\begin{eqnarray}
M \gg 1 \ .
\label{244}
\end{eqnarray}
(b) The energy shifts
$\tilde E_n-E_n$ must remain 
so small that $\tmax$ in 
(\ref{203}) is still much 
larger than
the actual relaxation time 
scale of the system under 
consideration.

Since generic level populations
$p_n$ and energy level distances
are extremely small 
(see beginning of this section) 
and in view of
the possibility to ``redistribute''
the $p_n$'s among the $\tilde p_n$'s
(see below  (\ref{201b}))
and to ``rearrange'' the energy levels
(see (\ref{203})), it seems
reasonable to expect that the above
described construction can be successfully
implemented in many cases of interest.
One particularly simple possibility
is as follows:

Assuming that the system exhibits a macroscopically 
well-defined energy (see above equation 
(\ref{20}) and beginning of this section), 
there exists a microcanonical energy window
$W:=[E-\Delta E, E]$, whose width 
$\Delta E$ is small on the macroscopic 
scale, but still so large that we can set 
$\tilde p_n=0$ for all $n$ with $E_n\not\in W$
(see below equation (\ref{201b})).
In other words, the set $I$ in (\ref{203a})
only contains $n$'s with $E_n\in W$.
Similarly as above (\ref{14}), we 
can and will temporally redefine the
corresponding indices so that 
$n\in\{1,...,D\}$ for all those $E_n$'s contained in 
$W$, and thus $I=\{1,...,D\}$.
Moreover, we can assume without loss
of generality that those $E_n$'s are 
ordered by magnitude (i.e. $E_{n+1}\geq E_n$
for all $n\in\{1,...,D-1\}$).
In a second step, we define $\tilde M$ as 
the smallest integer with the property that 
$\tilde M\geq1/\sqrt{\tilde\pmax}$,
where $\tilde\pmax:=\max_n \tilde p_n$.
According to the discussion at the
beginning of this section, $\tilde\pmax$
will usually be exponentially small in $f$
for a system with $f$ degrees of 
freedom, hence $\tilde M$ will be 
exponentially large in $f$.
Next, we choose $I_1:=\{1,...,D_1\}$,
where $D_1$ is the smallest integer 
with the property that 
$\sum_{n=1}^{D_1}\tilde p_n\geq 1/\tilde M$.
Finally, the latter inequality can be turned 
into an equality, i.e.,
$\sum_{n=1}^{D_1}\tilde p_n=1/\tilde M$,
by slightly reducing some 
of the $\tilde p_n$'s with $n\leq D_1$ 
(and at the same time slightly increasing 
some with $n>D_1$).
By modifying the $\tilde p_n$'s along this
line, one readily sees that the original sum 
on the left hand side of (\ref{201b}) 
may increases at most by $2\tilde\pmax$.
Likewise, $I_2:=\{D_1+1,...,D_2\}$,
where $D_2$ is the smallest integer 
with $\sum_{n=D_1+1}^{D_2}\tilde p_n\geq1/\tilde M$;
then the $\tilde p_n$'s are again slightly adjusted 
so that $\sum_{n=D_1+1}^{D_2}\tilde p_n=1/\tilde M$;
and so on for $I_3,...,I_{\tilde M}$.
Altogether, the original sum 
on the left hand side of (\ref{201b}) thus
may increases at most by $2\tilde\pmax \tilde M$,
which is still exponentially small in $f$.
In a third step, we define $D_0:=1$ and
$\delta E_\mu:=E_{D_\mu}-E_{D_{\mu-1}}$
for $\mu=1,...,\tilde M$, i.e., $\delta E_\mu$
quantifies the energy variations within 
the subset $I_\mu$.
Let us now focus on the set $S$ of all
$\mu$'s with the property that 
$\delta E_\mu>\Delta E/\sqrt{\tilde M}$.
Observing that $D_{\tilde M}=D$ and 
$\sum_{\mu=1}^{\tilde M}\delta E_\mu = E_D-E_1 \leq \Delta E$,
the number of elements contained in $S$,
henceforth denoted as $|S|$, must satisfy
$|S|\leq\sqrt{\tilde M}$.
In turn, the complement 
$\bar S:=\{1,...,\tilde M\}\setminus S$
contains $M:=\tilde M-|S|$ elements.
It readily follows that $M$ is still 
exponentially large in $f$.
The last step consist in redistributing the 
populations $\tilde p_n$ of all subsets 
$I_\mu$ with $\mu\in S$ uniformly among 
those with $\mu\in \bar S$.
By construction, after this redistribution of 
the $\tilde p_n$'s, the ``new'' value of
$\sum_{n\in I_\mu}\tilde p_n$ 
is thus equal to $1/M$ if 
$\mu\in\bar S$ and zero otherwise.
Furthermore, the contribution of this final redistribution 
of the $\tilde p_n$'s to the left hand side
in (\ref{201b}) can be upper bounded by
$2|S|/\tilde M$, which is still exponentially
small in $f$.
If we now change the labels $\mu$ so
that $\bar S=\{1,...,M\}$ and define
$E'_{\mu}:=E_{D_\mu}$, then all requirements
of our above described construction are fulfilled.
In particular, $\tmax$ in (\ref{203}) will be
exponentially large in $f$.

\subsection{Derivation of the main result}
\label{s53}
In order to explain the main ideas,
we temporarily focus on pure states
$\rho(t)$
(for mixed states see section \ref{s56}).
Hence, also $\tilde\rho (t)$ is
pure (see below (\ref{201b})), i.e., 
there exist certain (normalized) 
vectors $|\psi (t)\rangle$ 
and $|\tilde\psi (t)\rangle$
so that
\begin{eqnarray}
\rho(t) & = & |\psi(t)\rangle\langle\psi(t)| \ ,
\label{220a}
\\
\tilde \rho(t) & = & 
|\tilde \psi(t)\rangle\langle\tilde \psi(t)| \ .
\label{220}
\end{eqnarray}
Since the dynamics of $\rho(t)$ is governed by 
the Hamiltonian $H$ from (\ref{5})
and that of
$\tilde\rho(t)$ by $\tilde H$ 
from (\ref{204}), 
it follows that
\begin{eqnarray}
|\psi(t)\rangle & = & e^{-i Ht/\hbar}|\psi(0)\rangle
\ ,
\label{220z}
\\
|\tilde\psi(t)\rangle & = & 
e^{-i\tilde Ht/\hbar}|\tilde\psi(0)\rangle
\ ,
\label{220''}
\end{eqnarray}
see also (\ref{4}) and (\ref{6}).
Exploiting (\ref{220}), the level populations 
in (\ref{201a})
can be rewritten as
\begin{eqnarray}
\tilde p_n = |c_n|^2 \ ,
\label{220'}
\end{eqnarray}
where $c_n:=\langle n|\tilde\psi(0)\rangle$.
Since $\tilde p_n=0$ unless 
$n\in I$ (see (\ref{203a}))
it follows that
\begin{eqnarray}
|\tilde\psi(0)\rangle & = & \sum_{n\in I} c_n |n\rangle \ .
\label{221}
\end{eqnarray} 

In passing we note that 
a pure state like in (\ref{220a})
may still exhibit a small population 
$p_n$ of every single energy 
level, as required throughout 
our present approach.
In particular, the diagonal ensemble in  (\ref{15}),
which governs the long-time behavior
(after equilibration) will then exhibit
a small purity $\mbox{Tr}\{\rhobar^2\}$ notwithstanding 
the fact that we are dealing with a pure state,
i.e., $\mbox{Tr}\{[\rho(0)]^2\}=1$.

Taking for granted that 
the construction from the previous subsection
has been successfully implemented,
the approximation
\begin{eqnarray}
\langle A\rangle_{\!\rho(t)}=
\langle\tilde\psi(t)|A|\tilde\psi(t)\rangle
\label{245}
\end{eqnarray}
will thus be fulfilled very well
for all $t\ll\tmax$.
Furthermore, it follows from
(\ref{240}) and (\ref{220'})
that the vectors
\begin{eqnarray}
|\psi'_\mu\rangle:=\frac{1}{\sqrt{p_\mu'}}
\sum_{n\in I_\mu}c_n\, |n\rangle
\label{250}
\end{eqnarray}
satisfy
\begin{eqnarray}
\langle\psi'_\mu|\psi'_\nu\rangle=\delta_{\mu\nu}
\label{251}
\end{eqnarray}
and that (\ref{221})
can be rewritten as
\begin{eqnarray}
|\tilde\psi(0)\rangle = \sum_{\mu=1}^M \sqrt{p_\mu'}\, 
|\psi'_\mu\rangle
\ .
\label{260}
\end{eqnarray}
Moreover, we can infer from (\ref{204}) and
(\ref{241}) that
\begin{eqnarray}
\tilde H |\psi'_\mu\rangle = E'_\mu |\psi'_\mu\rangle
\label{270}
\end{eqnarray} 
and with (\ref{220''}) and (\ref{260}) that
\begin{eqnarray}
|\tilde\psi(t)\rangle= \sum_{\mu=1}^M \sqrt{p_\mu'}
\, e^{-iE'_\mu t/\hbar}\, |\psi'_\mu\rangle \ .
\label{280}
\end{eqnarray}
Exploiting (\ref{245}), we finally arrive at
\begin{eqnarray}
\langle A\rangle_{\!\rho(t)}
& = &
\sum_{\mu,\nu=1}^M 
e^{i(E'_\nu-E'_\mu)t/\hbar}\,
\rho'_{\mu\nu}(0) A'_{\nu\mu}
\ ,
\label{320}
\\
A'_{\nu\mu}
&:= &
\langle\psi'_\nu|A|\psi'_\mu\rangle
\ ,
\label{300}
\\
\rho'_{\mu\nu}(0)
&:=&
\langle\psi'_\mu|\tilde\rho(0)|\psi'_\nu\rangle
=
\sqrt{p'_\mu p'_\nu} \ ,
\label{310}
\end{eqnarray}
where the last relation follows 
from (\ref{220}) and (\ref{260}).
In particular, $\rho'_{\mu\nu}(0)$ is 
a well defined $M\times M$ density matrix 
(Hermitian, positive, of unit trace).

The right hand side of (\ref{320}) is 
formally identical to that of (\ref{17}).
But now all level populations are 
equal (see (\ref{243})),
i.e., we got rid of the shortcomings
mentioned at the beginning of 
section \ref{s5}.

At this point, the assumption (a) from 
(\ref{244}) is needed. Namely, due to
this assumption and the formal equivalence 
of (\ref{320}) with (\ref{17}),
the heuristic considerations
from section \ref{s2} or the more rigorous
treatment in  \cite{rei16,bal17} 
can be adopted to arrive at the 
counterpart of (\ref{90}), namely
\begin{eqnarray}
\langle A\rangle_{\!\rho(t)} 
& = & 
\langle A\rangle_{\!\rhobar' }
+
G(t)\, \left[\langle A\rangle_{\!\rho'\!(0)} 
- \langle A\rangle_{\!\rhobar' }\right]
\ ,
\label{321}
\\
G(t) & := & (M|\chi(t)|^2-1)/(M-1)
\ ,
\label{340}
\\
\chi(t) 
& := & 
\frac{1}{M}\sum_{\mu=1}^M e^{iE'_\mu t/\hbar}
\ ,
\label{350}
\\
\rho'(0)
& := &
\sum_{\mu,\nu=1}^M \sqrt{p'_{\mu}p'_\nu} \, |\psi'_\mu\rangle\langle\psi'_\nu|
\ ,
\label{331}
\\
\rhobar'
& := &
\sum_{\mu=1}^M p'_{\mu} \, |\psi'_\mu\rangle\langle\psi'_\mu|
\ .
\label{330}
\end{eqnarray}
Exploiting (\ref{244}) once more, 
one can infer from (\ref{340}),
similarly as in (\ref{100}),
the very accurate approximation
\begin{eqnarray}
G(t) & = & |\chi(t)|^2 \ .
\label{360}
\end{eqnarray}

Upon comparison of $\chi(t)$ in (\ref{350})
with $\phi(t)$ in (\ref{70}), the main
properties of $G(t)$ in (\ref{360}) 
readily follow from those of $F(t)$ 
in (\ref{100}), see above  (\ref{110}):
(i) $G(0)=1$.
(ii) $0\leq G(t)\leq 1$ 
for all $t$.
(iii) $G(t)$ remains negligibly 
small for the vast majority of 
all sufficiently large times $t$.
In the latter statement we took 
(\ref{244}) for granted and we assumed 
without loss of generality that the 
$E'_\mu$ in (\ref{241}) were chosen so that
$E'_\mu\not=E'_\nu$ for all $\mu\not=\nu$.

Setting $t=0$ in (\ref{321}), 
the above property (i) implies that
\begin{eqnarray}
\langle A\rangle_{\!\rho'\!(0)}
=
\langle A\rangle_{\!\rho(0)} \ .
\label{370}
\end{eqnarray}
More precisely, (\ref{370}) is an
approximation of the same quality 
as (\ref{321}) itself.
Next we make use of the assumption 
(b) below (\ref{244}) that 
$\langle A\rangle_{\!\rho(t)}$
approaches its approximately constant
long-time limit already 
for times $t$ much 
smaller than $\tmax$ in (\ref{203}).
On the one hand, for (most of)
those times $t$ the result 
(\ref{321}) is still valid and 
the function $G(t)$ therein  
must assume values close to zero.
On the other hand, we know
from section \ref{s1} that 
$\langle A\rangle_{\!\rho(t)}$
stays very close to
$\langle A\rangle_{\!\rhobar}$
for most $t$'s beyond the 
initial relaxation time span.
We thus can conclude that in very good 
approximation
\begin{eqnarray}
\langle A\rangle_{\!\rhobar'}
=
\langle A\rangle_{\!\rhobar} \ .
\label{372}
\end{eqnarray}

By introducing (\ref{370}) and (\ref{372})
into (\ref{321}) we arrive at the
main new result of our paper,
namely
\begin{eqnarray}
\langle A\rangle_{\!\rho(t)} 
& = & 
\langle A\rangle_{\!\rhobar}
+
G(t)\, \left[\langle A\rangle_{\!\rho(0)} 
- \langle A\rangle_{\!\rhobar}\right] \ .
\label{375}
\end{eqnarray}

\subsection{Discussion of $G(t)$}
\label{s54}
A first set of basic qualitative features 
of $G(t)$ are the properties 
(i)-(iii) mentioned below (\ref{360}).
The remainder of this subsection is devoted
to recasting $G(t)$ from (\ref{360}) 
and (\ref{350}) into physically
more illuminating and practically more
convenient forms.

By utilizing the approximation 
(\ref{243}) and the definition 
(\ref{240}) we can conclude with
(\ref{350}) that
\begin{eqnarray}
\chi(t)
=
\sum_{\mu=1}^M\sum_{n\in I_\mu}
\tilde p_n e^{i E'_\mu t/\hbar}
\ .
\label{430}
\end{eqnarray}
Observing (\ref{241}) and that
the set $I$ is the disjoint union of  
the subsets $I_1,..,I_M$ 
(see above (\ref{241})) implies
\begin{eqnarray}
\chi(t)
=
\sum_{n\in I}
\tilde p_n e^{i\tilde E_n t/\hbar}
\ .
\label{440a}
\end{eqnarray}
Since $\tilde p_n=0$ for $n\not\in I$
(see (\ref{203a})) we arrive at
\begin{eqnarray}
\chi(t)
& = &
\sum_{n}
p_n e^{i\tilde E_n t/\hbar} + \delta
\ ,
\label{440}
\\
\delta & := &
\sum_{n}
(\tilde p_n - p_n)\, e^{i\tilde E_n t/\hbar} 
\label{440b}
\ .
\end{eqnarray}
By similar (but simpler) calculations
as in Appendix B (especially around 
 (\ref{b300})) in combination
with our assumption (\ref{203})
one finds that the $\tilde E_n$'s 
in (\ref{440}) can be very well approximated 
by the $E_n$'s.
Furthermore, $\delta$ from (\ref{440b})
can be safely neglected in (\ref{440})
due to our assumption (\ref{201b}).
Exploiting  (\ref{201a}),
we thus obtain
as a first main result of this 
subsection 
\begin{eqnarray}
\chi(t) & = & \sum_{n}\rho_{nn}(0)\, e^{iE_nt/\hbar}
=\mbox{Tr}\{\rho(0)\, e^{iHt/\hbar}\}
\ .
\label{450}
\end{eqnarray}
This is the announced amendment
of (\ref{70}), quantitatively accounting
for our previous expectation that larger 
level populations $\rho_{nn}(0)$
should somehow play
a more important role than smaller 
ones.

Next we rewrite (\ref{450}) in the equivalent form
\begin{eqnarray}
\chi(t) & = & \int dE\ \rho(E)\ e^{iEt/\hbar}
\ ,
\label{451}
\\
\rho(E)
& := &
\sum_n \rho_{nn}(0)\ \delta(E-E_n) \ .
\label{452}
\end{eqnarray}
The function $\rho(E)$ thus quantifies the 
detailed population of all the energy levels, 
and $\chi(t)$ is its Fourier 
transform\footnote{Likewise,
$G(t)$ in (\ref{360}) may be 
viewed as the Fourier transform of
$\rho_2(E):=\int dE'\, \rho(E-E')\,\rho(E')$
(self-convolution of $\rho(E)$).}.
Usually, the energies $E_n$ are extremely 
dense and the sum of delta functions 
in (\ref{452}) can 
be replaced by a reasonably smoothened
approximation without any notable change
of $\chi(t)$ in (\ref{451}) during the 
entire initial relaxation time period, 
see also Appendices A and B.
In other words, $\rho(E)$
may be viewed as the smoothened 
(coarse grained) energy
distribution of the system.
While this distribution is hardly ever
available in experiments, it often is
in numerical simulations, as 
exemplified in section \ref{s7}.

The same approximation as for $F(t)$ in (\ref{110})
is readily recovered for $G(t)$ via 
(\ref{360}) and (\ref{451}) 
if the $\rho_{nn}(0)$ in (\ref{452})
are (approximately) 
equally large for all $E_n$ below some 
threshold energy $E$ and (practically)
negligible for all $E_n>E$, and
provided that the Hamiltonian $H$ exhibits 
reasonable thermodynamic properties
(well defined entropy $S(E)$ and 
(positive, intensive) temperature 
$T:=1/S'(E)$).
The same result 
\begin{eqnarray}
G(t)=1/[1+(t\,\kB T/\hbar)^2] 
\label{110a}
\end{eqnarray}
still applies if only
energies $E_n$ within a microcanonical 
energy window $[E-\Delta E,E]$ contribute,
as long as its width $\Delta E$ is much
larger than the thermal energy $\kB T$,
as it is usually the case.
More precisely, it is only the coarse 
grained $\rho(E)$ (see below (\ref{452}))
that must closely resemble the one
which would be obtained for strictly equally
large $\rho_{nn}(0)$'s for all
$E_n\in [E-\Delta E,E]$.
The actual $\rho_{nn}(0)$'s 
(before coarse graining) may 
thus still exhibit quite considerable
``fine grained'' variations.
In other words, the approximation (\ref{110a})
is found to remain valid under substantially
weaker premises than its predecessor
in  (\ref{110}).

Instead of such a microcanonical distribution,
one might also consider a canonical distribution,
i.e., the $\rho_{nn}(0)$'s are (approximately)
proportional to $\exp\{-E_n/k_B T\}$.
Similarly as in (\ref{110}), a 
straightforward calculation then yields
\begin{eqnarray}
G(t)=\exp\{-(t\, k_B T/\hh)^2 dE(T)/d(k_B T)\}
\ .
\label{455}
\end{eqnarray}
Note that $dE(T)/d T$ is the system's 
specific heat and $dE(T)/d(k_B T)$ is a 
dimensionless number which is typically 
comparable in order of magnitude
to the number $f$ of the system's 
degrees of freedom.
However, it must be emphasized that
there is no reasonable argument of 
why the far from equilibrium 
initial state $\rho(0)$ at time $t=0$ 
should exhibit a canonical energy 
distribution in the basis of the Hamiltonian 
$H$ which governs the relaxation dynamics 
of the isolated system for $t>0$.

For systems {\em at thermal equilibrium}, 
the so-called equivalence of ensembles 
is often taken for granted under quite 
general conditions.
However, no such equivalence 
is to be expected for the temporal
relaxation of {\em far from equilibrium}
initial states, as exemplified by
the very different findings
(\ref{110a}) and (\ref{455}).

More generally speaking, the above 
examples illustrate the fact that the 
function $G(t)$ depends on the details 
of the initial energy distribution, but 
does not depends on any further 
properties of the initial condition.

Taking into account (\ref{5}), (\ref{220a}), and
(\ref{220z}), one can rewrite (\ref{450}) as
\begin{eqnarray}
\chi(t)=\langle \psi(t)|\psi(0)\rangle \ ,
\label{460}
\end{eqnarray}
i.e., $\chi(t)$ represents the overlap
between the time evolved state and
the initial state.
Similarly, (\ref{360}) takes the form
\begin{eqnarray}
G(t)=|\langle \psi(t)|\psi(0)\rangle|^2 \ ,
\label{470}
\end{eqnarray}
i.e., $G(t)$ may be viewed as a survival 
probability (of the initial state) or
return probability (of the time evolved 
state), sometimes also denoted as 
(quantum) fidelity.

Mathematically speaking,  (\ref{220z})
and (\ref{460}) immediately imply that
\begin{eqnarray}
\chi(t)=\langle \psi(t+s)|\psi(s)\rangle
\label{480}
\end{eqnarray}
for any, arbitrary but fixed reference time point 
$s\in\RR$.
Physically speaking, this observation is quite
remarkable:
The crucial function $G(t)$ in (\ref{375})
can be recovered from the overlap decay
in (\ref{480}) with respect to any 
time evolved state $|\psi(s)\rangle$ of the system,
even if the reference time $s$ is chosen
very ``late'' and thus one might have expected that
the system has already equilibrated in any 
meaningful sense, and, in particular, has
``forgotten'' the initial disequilibrium conditions.

\subsection{Summary and discussion}
\label{s55}
The main result of this section consists 
in the approximation (\ref{375}) 
for the temporal relaxation,
where $G(t)$ in (\ref{360})
follows from either of the 
equivalent forms (\ref{450}), 
(\ref{451}), or (\ref{460}).
They encapsulate the details of how
the function $G(t)$ in (\ref{375})
decays from its initial value $G(0)=1$
towards $G(t)\simeq 0$ for (most)
sufficiently large $t$.
In particular, upon rewriting 
(\ref{375}) as 
\begin{eqnarray}
\frac{\langle A\rangle_{\!\rho(t)} 
-
\langle A\rangle_{\!\rhobar}}
{\langle A\rangle_{\!\rho(0)} 
-
\langle A\rangle_{\!\rhobar}}
=G(t)
\ ,
\label{375a}
\end{eqnarray}
taking for granted the assumptions underlying 
this result (see below),
and observing that $G(t)$
in (\ref{470})
is independent of $A$, 
we can conclude that, for any given $\rho(0)$,
the left hand side in (\ref{375a}) exhibits 
for all observables $A$ the same temporal 
relaxation behavior.

Provided that the additional information 
required in (\ref{450}), 
(\ref{451}), or (\ref{460}) is available,
this result (\ref{375}) represents
a significant step beyond the
previously known approximation
(\ref{90}), wherein $F(t)$ follows
from (\ref{70}) and (\ref{100}).

In particular, 
to determine $F(t)$ one usually needs to 
explicitly specify some appropriate energy window
(see above equation (\ref{20})).
In addition, in order to evaluate (\ref{70}) 
and (\ref{100}), one must determine the
eigenvalues of the Hamiltonian.
In contrast, $G(t)$ can be determined
via (\ref{470}) without explicitly specifying
some energy window and without 
diagonalizing the Hamiltonian.

The underlying key idea and main 
requirements essentially
amount to the following three steps: 
To begin with, all extremely small
level populations $p_n$ 
are neglected.
The remaining, non-negligible $p_n$'s are
then distributed into subsets $I_\mu$
with approximately equal net
populations $\sum_{n\in I_\mu}p_n$.
Moreover, all energies 
$E_n$ belonging to the same 
subset must be very close 
to each other.
In the end, the initially neglected
$p_n$'s are redistributed among the
subsets, and also the non-negligible
$p_n$'s may still be slightly adjusted,
the main aim being to further equalize
the subset populations.

Once such a rearrangement of the 
energy eigenvalues and redistribution of 
the level populations is accomplished, 
the same arguments as in section \ref{s2} 
or in \cite{rei16,bal17} can be adopted to 
arrive at (\ref{375}).
In so far as these arguments are 
non-rigorous (no error bounds or 
systematic improvements or are 
available), the result  (\ref{375})
may be viewed as an approximative 
proposition of the same character.

The remaining requirements are largely
the same as in sections \ref{s3} and \ref{s4}.
The basic reason is that the prediction
(\ref{375}) is essentially a 
modification of  (\ref{90}), it is not 
expected to cover previously excluded cases.

In passing we note that when focusing for a 
given pure state (\ref{220a}) on the 
particular observable 
$A=|\psi(0)\rangle\langle\psi(0)|$, 
then the expectation value on the left hand 
side of (\ref{375}) coincides exactly 
with the survival probability in (\ref{470}).
On the right hand side of (\ref{375}),
one readily find that 
$\langle A\rangle_{\!\rho(0)}=1$
and $\langle A\rangle_{\!\rhobar}
=\sum_n p_n^2\leq \max_np_n$.
Since $p_n\ll 1$ for all $n$ 
(see (\ref{16}) and beginning of section \ref{s5}),
our result (\ref{375}) thus reproduces
the exact result very well in this 
special case.
The latter exact result apparently goes 
back to Torres-Herrera, Vyas, and Santos
(see \cite{tor14,tor15} and further 
references therein), hence our 
present work may be viewed as a 
generalization of theirs.

\subsection{Mixed states}
\label{s56}
So far, our main result (\ref{375}) 
has only be justified for pure states 
(see section \ref{s53}).
Turning to mixed states, we recall that
any given density operator $\rho$ can 
be written in the form
\begin{eqnarray}
\rho=\sum_{j=1}^J w_j\,|\psi_j\rangle\langle \psi_j |
\label{500}
\end{eqnarray}
for some suitably chosen set of pure (normalized) 
states $|\psi_j\rangle$ and weights $w_j\geq 0$
with $\sum_{j=1}^J w_j=1$.
In general, the vectors $|\psi_j\rangle$ need not
be pairwise orthogonal and not even linearly 
independent, hence there usually exist many 
different ``representations'' (\ref{500}) 
of the same density operator $\rho$.
The same properties remain true when
the density operator and the pure 
states in (\ref{500})
acquire a time dependence
via the pertinent Liouville-von Neumann
and Schr\"odinger equations, respectively.
Such a time dependence is henceforth 
tacitly assumed in (\ref{500}),
while arguments $t$ are still omitted.

Taking for granted that every pure state
$|\psi_j\rangle$ in (\ref{500}) 
satisfies the requirements from section \ref{s55},
the approximation (\ref{375}) will be valid for
each of them.
Next we observe that all expectation values 
appearing in (\ref{375}) are linear functionals 
of $\rho$.
But in general, also $G(t)$ on the right 
hand side is a non-trivial (non-linear) 
functional of $\rho$ according to (\ref{360}) 
and (\ref{450}).
It follows that (\ref{375}) cannot be 
valid in full generality
(the left hand side is linear and the right 
hand side non-linear in $\rho$).
However, under the extra assumption that 
$G(t)$ is (approximately) identical for all 
$|\psi_j\rangle$ with non-negligible weights 
$w_j$ in (\ref{500}), 
one readily concludes that 
also their linear combination in (\ref{500})
will satisfy (\ref{375}),
where the symbols $\rho$ and
$\rhobar$ in (\ref{375})
now refer to the actual density 
operator $\rho$ on the
left hand side of (\ref{500}),
and likewise for the $\rho$'s
appearing in (\ref{450})-(\ref{452}).
It seems reasonable to expect that
such approximately identical $G(t)$'s
may arise -- at least for one of the many
possible representations (\ref{500})
of the same $\rho$ -- in many cases
of interest.

In fact, if the initial state 
$\rho(0)$ is of low purity 
(``strongly mixed''),
i.e.,
$\mbox{Tr}\{[\rho(0)]^2\}\ll 1$,
it is rigorously shown in 
Appendix C that our main result 
(\ref{375}) still amounts
to a very good approximation,
where $G(t)$ is again given by
(\ref{360}) and (\ref{450}).
In other words, (\ref{375}) is
known to apply both for pure and
strongly mixed states. 
Once again, it is therefore
quite plausible that
the same result will remain
(approximately) correct 
also in the intermediate 
case, i.e., when the purity 
$\mbox{Tr}\{[\rho(0)]^2\}$
is neither unity nor 
close to zero, 
see also end of Appendix C.
However, providing a more rigorous 
demonstration or criterion appears 
to be a very daunting task.

\section{Comparison with numerics}
\label{s7}
As already mentioned at the beginning 
of section \ref{s6}, the spatial homogeneity 
requirement of our present theory 
considerably restricts the number of 
suitable experimental and numerical 
examples in the literature, with which
it might be compared.
Moreover, our amended
theoretical prediction (\ref{375}) 
requires information about 
the function $G(t)$ in (\ref{360})
and thus either about the level populations 
in (\ref{450})-(\ref{452}) or about the overlaps 
in (\ref{460}), which is not available 
in most experiments up to now.
However, it is noteworthy that
the overlap of two quantum many-body 
states has recently been successfully 
measured for ultra-cold bosonic atoms 
in optical lattices \cite{isa15},
hence a direct comparison of our
theory with experiments may 
become feasible in the future.
With respect to numerical results,
the latter information should in principle
be accessible quite often, but in
practice it is provided as published 
data in a relatively small number 
of cases.
In the following, we compare
our theory with two such examples, 
for which all the
necessary data are available.

\begin{figure}
\epsfxsize=1.0\columnwidth
\epsfbox{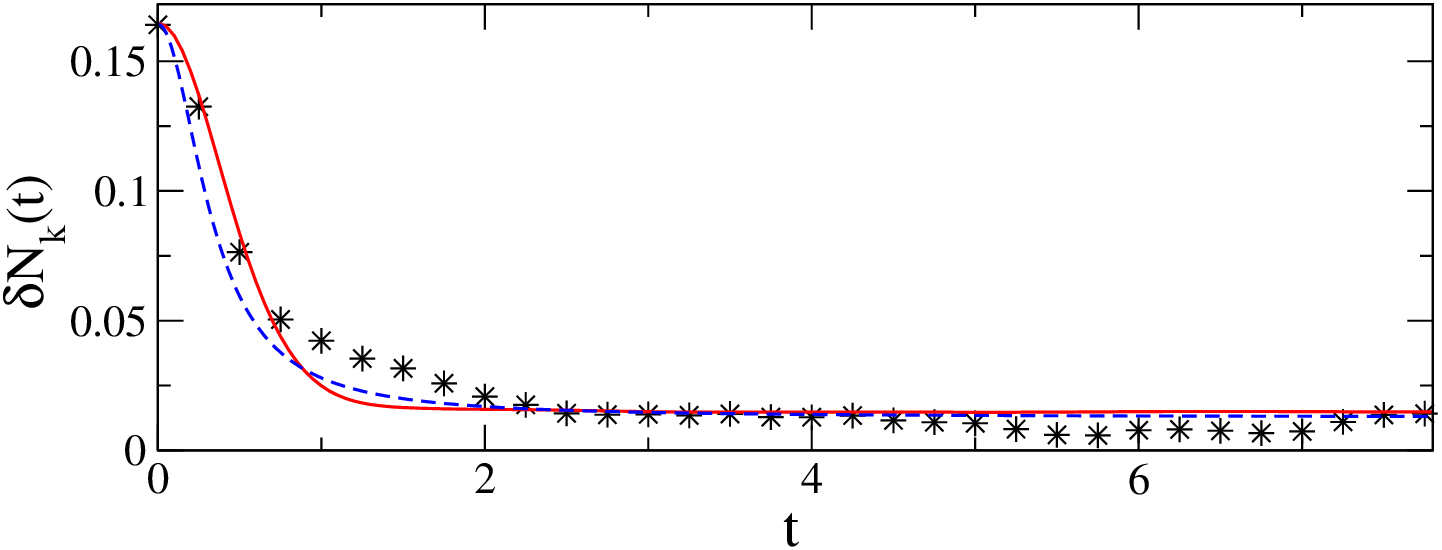}
\caption{\label{fig2}
Symbols: 
Numerical data from figure 2(g) 
in \cite{rig09} for the density-density 
structure factor $\delta N_k(t)$
of a one-dimensional fermionic
model system
(for more details see main 
text and  \cite{rig09}).
Solid: 
Theoretical prediction 
from (\ref{375}), where $G(t)$ 
was evaluated according to 
(\ref{360}) and (\ref{450})
by employing the numerically 
determined values of $E_n$ and 
$\rho_{nn}(0)$ from 
 \cite{rig09}, 
see figure 7(a) therein
(the original data were 
kindly provided by Marcos 
Rigol).
Dashed: 
Theoretical prediction from 
(\ref{90}) and (\ref{110})
(or from (\ref{375}) and (\ref{110a})),
adopting the estimate $T=3$ 
provided by  \cite{rig09}.
Both in (\ref{90}) and (\ref{360}),
the quantitative values
of $\langle A\rangle_{\!\rho(0)}$ 
and $\langle A\rangle_{\!\rhobar}$
have been fitted to the numerical 
data.
Following  \cite{rig09}, the units 
have been chosen so that $k_B=\hbar=1$.
}
\end{figure}

Our first example is the extended Hubbard model
for 8 strongly correlated fermions on a one 
dimensional lattice with 24 sites, 
whose thermalization after a quantum
quench has been numerically
explored by Rigol in  \cite{rig09}.
figure \ref{fig2}  exemplifies
a representative non-integrable 
case with nearest-neighbor hopping
and interaction parameters 
$\tau=V=1$ and next-nearest-neighbor 
hopping and interaction parameters 
$\tau'=V'=0.32$, corresponding to 
the data from
figures 2(g) and 7(a) in \cite{rig09}.
The numerical findings are
compared in figure \ref{fig2}
with the amended theory from
(\ref{360}), (\ref{375}), and (\ref{450}),
as well as with its predecessor from 
(\ref{90}) and (\ref{110}),
or, equivalently, the approximation 
from (\ref{375}) and (\ref{110a}).
In view of the still quite notable 
numerical finite size fluctuations 
(8 Bosons on 24 sites),
whose magnitude can be estimated
from the non-stationarity of the
numerical data beyond the actual 
relaxation time span in figure 
\ref{fig2} (see also figure 2(g) 
in  \cite{rig09}), it is 
impossible to decide which of 
the two theoretical curves 
exhibit a better agreement.
Within these numerical finite size
effects (which are beyond the 
theory) both curves agree reasonably 
well with the data.
We also may recall that the only 
fit parameters of the theory are
 the initial value $\langle A\rangle_{\!\rho(0)}$ 
and the long-time average 
$\langle A\rangle_{\!\rhobar}$.
As already mentioned in section \ref{s6},
the quantitative determination of those
two values for the quite elaborate observable
at hand (a dimensionless descendant of
the density-density structure factor \cite{rig09}) 
is not a main objective of our present work.

\begin{figure}
\epsfxsize=0.95\columnwidth
\epsfbox{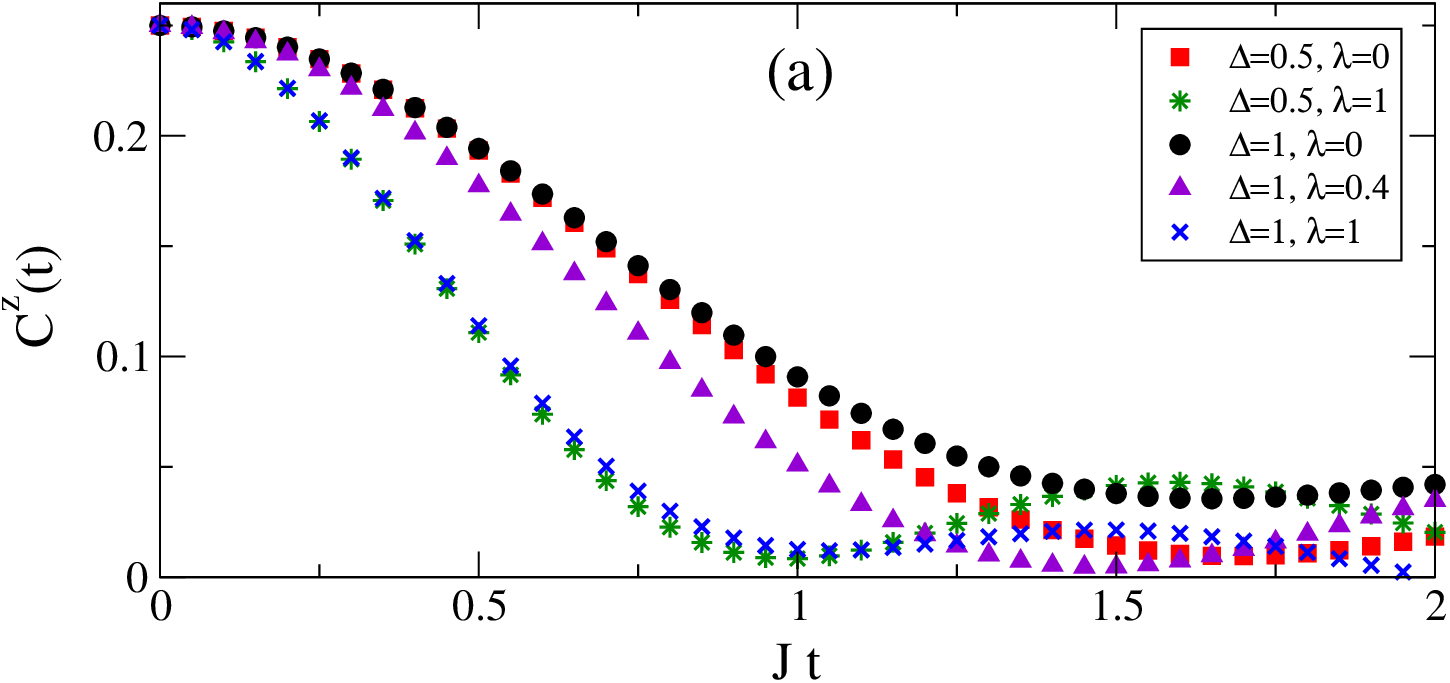}
\epsfxsize=0.95\columnwidth
\epsfbox{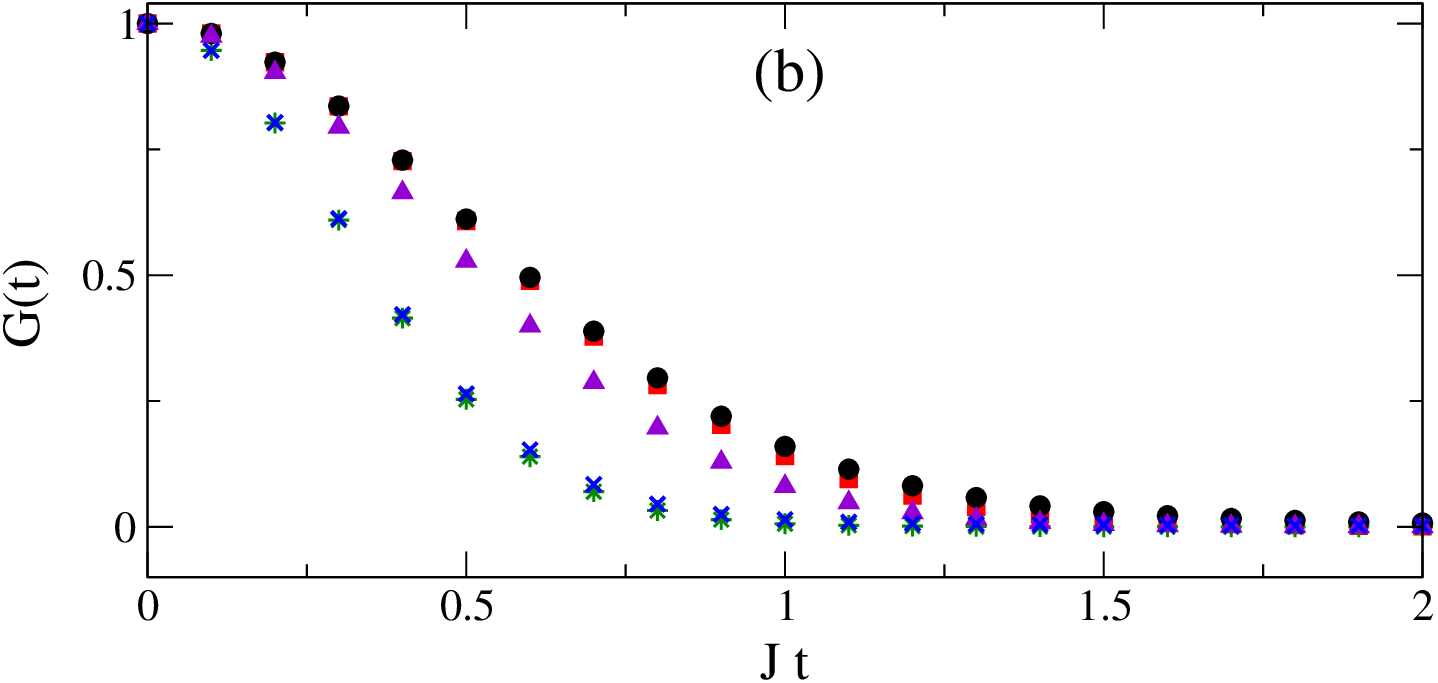}
\caption{\label{fig3}
(a)
Numerical results for the spin-spin correlation
$C^z(t)$, adopted from the top right plot in 
figure 8 of  \cite{tor14}.
The considered system is a one dimensional
spin-1/2 model with 16 spins, coupling J, 
anisotropy parameter $\Delta$, ratio
between nearest-neighbor and 
next-nearest-neighbor coupling 
$\lambda$, and $\hbar=1$.
The system is isotropic for 
$\Delta=1$ and anisotropic 
otherwise.
The case $\Delta\not=0$, $\lambda=0$ 
corresponds to the integrable XXZ model,
while the model is non-integrable
for $\Delta\not=0$, $\lambda\not=0$.
(b)
The corresponding numerical results for 
the survival probability $G(t)$ in (\ref{470}),
adopted from the top right plot 
in figure 5 of  \cite{tor14}.
}
\end{figure}

Our second example is the spin-chain model,
numerically explored by Torres-Herrera, Vyas, 
and Santos in  \cite{tor14},
see figure \ref{fig3}.
Specifically, the relaxation of an initial state,
consisting of 8 alternating pairs of parallel spins
is observed via the correlation $C^z(t)$ of two 
neighboring spins in the middle of the chain
\cite{tor14}, for which the initial expectation 
value is known to be $C^z(0)=0.25$.
The two examples in figure \ref{fig3}
with $\lambda=0$ correspond
to integrable systems, which are in general
not expected to thermalize in the long-time
limit, while the three examples 
with $\lambda\not =0$
are non-integrable, hence $C^z(t)$ should
approach the thermal long-time limit zero.
This expected long-time behavior is reasonably 
but not extremely well fulfilled by the 
numerical results for the two integrable
and the three non-integrable
cases in figure \ref{fig3}(a).
In fact, temporal ``oscillations'' 
comparable to those of the cross- and 
star-symbols
in figure \ref{fig3}(a) for $t\in[1,2]$ 
are found to persist in all five cases
up to (practically) arbitrarily large 
times $t$ (not shown).
Similarly as in the previous example 
in figure \ref{fig2},
these persistent oscillations are probably 
due to the still relatively small system 
size (16 spins).
In other words, it seems reasonable 
to expect that the behavior of $C^z(t)$
for much larger systems may still 
deviate by $0.05$ (or even more)
from the corresponding results
in figure \ref{fig3}(a).
Analogously, the numerically obtained
results from \cite{tor14} for the survival 
probability $G(t)$ in (\ref{470})
are reproduced in figure \ref{fig3}(b).
Apparently, the numerical
finite size effects
for this quantity $G(t)$ are considerably
weaker than for the quantity $C^z(t)$
depicted in figure \ref{fig3}(a).

To connect these numerical results 
with our present theory,
$C^z(t)$ in figure \ref{fig3}(a)
must identified with 
$\langle A\rangle_{\!\rho(t)}$ 
in equation (\ref{375}),
while $G(t)$ in figure \ref{fig3}(b)
coincides with $G(t)$ in (\ref{375}).
Still, the theory does not imply
any prediction regarding either
of these two quantities themselves.
Rather, it predicts that the two
quantities should be related
to each other according to 
(\ref{375}).
In doing so, the initial value
$\langle A\rangle_{\!\rho(0)}$ 
appearing in (\ref{375})
is known to be 
$C^z(0)=0.25$ (see above).
Moreover, the long-time limit
$\langle A\rangle_{\!\rhobar}$
appearing in (\ref{375}) must
be estimated from the long-time
behavior of $C^z(t)$ in figure
\ref{fig3}(a).
In view of the above mentioned
finite-size effects of the numerical
data for $C^z(t)$ in figure 
\ref{fig3}(a),
the agreement between this theoretical
prediction of equation (\ref{375}) 
and the numerical findings in 
figure \ref{fig3} is quite satisfying.

\section{Conclusions}
\label{s10}
The main result of this paper is the following
approximation for the temporal relaxation of a 
(pure or mixed) state $\rho(t)$, whose dynamics
is governed by a Hamiltonian with energy eigenvalues 
$E_n$ and eigenstates $|n\rangle$:
\begin{eqnarray}
\mbox{Tr}\{\rho(t)A\} = \mbox{Tr}\{\rho(0)A\} \left|\sum_n
\langle n|\rho(0)|n\rangle\, e^{iE_nt/\hbar}\right|^2
\ ,
\label{600}
\end{eqnarray}
where the observable $A$ has been tacitly 
``rescaled'' so that the long-time average of the 
left hand side is zero.

The first main prerequisite for (\ref{600})
is that the system must
equilibrate at all, i.e., the left hand side
of (\ref{600}) must remain very close to a 
constant value (here assumed to be zero) 
for the vast majority of all sufficiently large
times $t$, where 
``very close'' is meant in comparison with
the full range of possible measurement 
outcomes of $A$.
To guarantee the latter equilibration
property,
we have taken for granted a set of
sufficient conditions, 
which are already rather weak,
and which could still be 
considerably weakened in principle.
Most importantly, it is required that
there are no degenerate energies and 
energy gaps
(i.e. the energy differences $E_m-E_n$
are non-zero and mutually different for 
all pairs $m\not=n$),
and that all level populations 
$\langle n|\rho(0)|n\rangle$ 
are small
(cf.  (\ref{11}) and (\ref{16})).
On the other hand, it is {\em not} 
required that the system exhibits 
thermalization, i.e., the
long-time average in (\ref{600})
may still be different from
the pertinent thermal equilibrium 
value.

The second main prerequisite for (\ref{600})
is the absence of any notable macroscopic 
transport currents,
caused, e.g.,  
by some initially unbalanced 
local densities.
Such a transportless relaxation
can usually be taken for granted if both the
system Hamiltonian and the initial state are
spatially homogeneous on macroscopic 
scales.
A more detailed discussion of further 
possible prerequisites for (\ref{600})
is provided by sections \ref{s3} and \ref{s4}
(see also sections \ref{s52} and 
\ref{s56}).
In fact, formulating conditions, which
are strictly sufficient for (\ref{600})
but not too restrictive for practical 
purposes, remains an open problem.
In this respect, the situation is somewhat
similar as in density functional
theory, random matrix theory, and other 
``non-systematic'', but practically very 
successful approximations.

The most striking property of  (\ref{600}) is that
the considered observable $A$ does not matter
in the last factor, which encapsulates the entire time
dependence of the relaxation.
Generically, this factor is unity for $t=0$ and
very close to zero for practically
all sufficiently late times.
Specifically for a pure initial state $|\psi(0)\rangle$,
the last factor in (\ref{600}) can be identified
with $|\langle\psi(t)|\psi(0)\rangle|^2$
(survival probability).
On the one hand, (\ref{600}) may thus be viewed
as a (very substantial) generalization of 
previous results by Torres-Herrera, Vyas, 
and Santos \cite{tor14,tor15}.
On the other hand, also the earlier results
from  \cite{rei16,bal17} are recovered
as a special case, namely when all
level populations $\langle n|\rho(0)|n\rangle$
can be approximated as being either strictly 
zero, or equal to some (small but finite) 
constant value.

In many cases of practical interest,
the last factor in (\ref{600}) can be further
approximated as $1/[1+(t \, \kB T/\hbar)^2]$, where
$T$ is the temperature after thermalization,
or, if the system does not thermalize,
the temperature of a thermalized auxiliary 
system with the same (macroscopic) energy
as the true system.
In general, transportless relaxation is thus 
predicted to be non-exponential in time,
and the relevant time scale $\hbar/\kB T$ 
to be very small.

In principle, all these predictions may be viewed 
as approximative propositions due to the 
non-rigorous line of reasoning adopted 
in section \ref{s2} or in \cite{rei16,bal17}.
On the other hand, they have been validated
by showing that they compare very favorably with 
various previously 
published experimental and numerical results
for systems, which satisfy
the above mentioned main prerequisites 
of the theory reasonably well.

\ack
Inspiring discussions with Ben N. Balz
and Lennart Dabelow are gratefully 
acknowledged.
Special thanks is due to Martin Eckstein
for pointing out a misconception in
a preliminary attempt to model
the experiment from  \cite{gie15},
and to Lea Santos and Marcos Rigol for
providing the raw data of their previously
published results in  \cite{tor14} and
\cite{rig09}, respectively.
This work was supported by the 
Deutsche Forschungsgemeinschaft 
(DFG) within the Research Unit FOR 2692
under Grant No. 397303734.
The author acknowledges support for the Article
Processing Charge by the Deutsche 
Forschungsgemeinschaft and the Open Access
Publication Fund of Bielefeld University.

\appendix
\setcounter{section}{1}
\section*{Appendix A}
A first main goal of this appendix is
to justify the approximation
at the beginning of section \ref{s2}
and the closely related 
approximation (\ref{201}) 
under the condition (\ref{201b}).
More precisely,
we consider an arbitrary but 
fixed density operator $\rho(0)$ 
with level populations $p_n=\rho_{nn}(0)$ 
(see (\ref{11}))
and an arbitrary but fixed set of 
``auxiliary populations'' $\tilde p_n$, 
satisfying $\tilde p_n\geq 0$ 
and $\sum_n\tilde p_n=1$.
The above mentioned main goal now
consist in demonstrating that there exists 
an ``auxiliary density operator'' 
$\tilde \rho(0)$
with level populations 
$\tilde \rho_{nn}(0)=\tilde p_n$
and with the property that
\begin{eqnarray}
|\langle A\rangle_{\!\rho(t)} - 
\langle A\rangle_{\!\tilde\rho(t)}|
& \leq &
\da\sqrt{\sum_n|p_n-\tilde p_n|}
\label{a10}
\end{eqnarray}
for arbitrary $t$ and $A$,
where the time evolution of
both $\rho(0)$ and $\tilde\rho(0)$ 
is governed by the Hamiltonian 
(\ref{5}), and where $\da$ is the 
range of the observable $A$, i.e., the 
difference between its largest and 
smallest eigenvalues.
Since any real measurement device
corresponding to the observable
$A$ has a finite range $\da$ as well
as a finite resolution $\dda$
(see also section \ref{s1}),
it follows that the two expectation values
on the left hand side of (\ref{a10})
are experimentally
indistinguishable if the sum on the 
right hand side is smaller than
$(\dda/\da)^2$.
Altogether, this amounts to the precise 
quantitative justification of the 
two above mentioned approximations.

A secondary goal of this appendix is 
to show that whenever $\rho(t)$ is a pure
state then $\tilde\rho (t)$ 
will be pure as well.

To begin with, we recall from 
the beginning of section \ref{s1}
the relations
\begin{eqnarray}
\langle A\rangle_{\!\rho(t)} 
& = & 
\mbox{Tr}\{\rho(t)A\}
\ ,
\label{a20}
\\
\rho(t) 
& = & 
\propa_t \rho(0) \propa_t^\dagger
\ ,
\label{a30}
\\
\propa_t 
& := &
e^{-iHt/\hh} \ .
\label{a40}
\end{eqnarray}
The left hand side of (\ref{a40}) 
is understood as usual:
\begin{eqnarray}
e^{-iHt/\hh} := \sum_n e^{-iE_nt/\hh}\,|n\rangle\langle n|
\ .
\label{a45}
\end{eqnarray}
Exploiting the cyclic invariance of
the trace in (\ref{a20}), we can conclude that
\begin{eqnarray}
\langle A\rangle_{\!\rho(t)} 
& = & 
\mbox{Tr}\{\rho(0)B\}
\ ,
\label{a50}
\\
B
& := & 
\propa_t^\dagger A \propa_t \ .
\label{a60}
\end{eqnarray}
For notational simplicity, the dependence 
of $B$ in (\ref{a60})
on $t$ has been omitted.

Focusing temporarily on the case 
that $p_n:=\rho_{nn}(0)>0$ for all $n$
we define
\begin{eqnarray}
g_n & := & \sqrt{\tilde p_n/p_n}
\ ,
\label{a70}
\\
P & := & \sum_n g_n\, |n\rangle\langle n|
\ ,
\label{a80}
\\
Q & := & \id-P = \sum_n (1-g_n)\, |n\rangle\langle n|
\ ,
\label{a90}
\\
\tilde \rho(0) & := & P\rho(0)P 
\ .
\label{a100}
\end{eqnarray}
From these definitions it follows that
\begin{eqnarray}
\tilde\rho_{nn}(0)
=
\langle n|P\rho(0)P|n\rangle
=
g_n^2\rho_{nn}(0)
=
\tilde p_n \ .
\label{a110}
\end{eqnarray}
In other words, $\tilde\rho(0)$ indeed
exhibits the given  level 
populations $\tilde p_n$.
Moreover, one readily verifies that $\tilde\rho(0)$ 
is a non-negative, Hermitian operator 
of unit trace, i.e., a well-defined 
density operator.

If $\rho(0)$ is a pure state, it can be
written in the form $|\varphi\rangle\langle\varphi|$
for some $|\varphi\rangle$ of the form
$\sum c_n\,|n\rangle$.
By means of (\ref{a80}) and (\ref{a100})
it follows that $\tilde\rho(0)$
can be rewritten as  $|\tilde\varphi\rangle\langle\tilde\varphi|$
with $|\tilde\varphi\rangle:=\sum g_n c_n\,|n\rangle$,
i.e. also $\tilde\rho(0)$ is a pure state.

Since the dynamics of $\tilde\rho(0)$ 
and of $\rho(0)$ are governed 
by the same Hamiltonian $H$, it
follows exactly as in (\ref{a20})-(\ref{a60})
that
\begin{eqnarray}
\langle A\rangle_{\!\tilde\rho(t)} 
& = & 
\mbox{Tr}\{\tilde\rho(0)B\}
\ .
\label{a120}
\end{eqnarray}
According to (\ref{a90}) we have $Q+P=\id$ and hence
\begin{eqnarray}
\mbox{Tr}\{\rho(0)B\} & = &
\mbox{Tr}\{Q\rho(0)B\}+\mbox{Tr}\{P\rho(0)B\} 
\ ,
\label{a130}
\\
\mbox{Tr}\{P\rho(0)B\} & = &
\mbox{Tr}\{P\rho(0)QB\}+\mbox{Tr}\{P\rho(0)PB\} 
\, . \ \ \
\label{a140}
\end{eqnarray}
Due to (\ref{a100}), the last term in (\ref{a140})
is equal to $\mbox{Tr}\{\tilde\rho(0)B\}$.
Together with (\ref{a50}) and (\ref{a120})
we thus can conclude that
\begin{eqnarray}
\Delta 
& := &  
\langle A\rangle_{\!\rho(t)} 
- \langle A\rangle_{\!\tilde\rho(t)} 
= R_1+R_2
\ ,
\label{a150}
\\
R_1 & := & \mbox{Tr}\{Q\rho(0)B\}
\ ,
\label{a160}
\\
R_2 & := & \mbox{Tr}\{P\rho(0)QB\} \ .
\label{a170}
\end{eqnarray}

Since $\rho(0)$ is a non-negative
Hermitian operator, there exists a Hermitian
operator $\sigma$ with the property that 
$\sigma^2=\rho(0)$.
Considering $\mbox{Tr}\{C_1^\dagger C_2\}$
as a scalar product between two arbitrary
linear (but not necessarily Hermitian) 
operators $C_{1,2}$, the Cauchy-Schwarz
inequality takes the form
$|\mbox{Tr}\{C_1^\dagger C_2\}|^2\leq
\mbox{Tr}\{C_1^\dagger C_1\}\mbox{Tr}\{C_2^\dagger C_2\}$.
Choosing $C_1=(Q\sigma)^\dagger$ and $C_2=\sigma B$
we can infer from (\ref{a160}) that
\begin{eqnarray}
|R_1|^2 
\leq 
\mbox{Tr}\{Q \sigma\sigma^\dagger Q^\dagger\}
\,
\mbox{Tr}\{B^\dagger \sigma^\dagger\sigma B\} \ .
\label{a190}
\end{eqnarray}
Observing that all operators on the right 
hand side of (\ref{a190}) are Hermitian and 
exploiting the cyclic invariance of the trace 
yields
\begin{eqnarray}
|R_1|^2 
\leq 
\mbox{Tr}\{\rho(0)\, Q^2\}
\,
\mbox{Tr}\{\rho(0)\, B^2\} \ .
\label{a200}
\end{eqnarray}
Evaluating the trace by means of 
the eigenbasis of $B$ results in
\begin{eqnarray}
\mbox{Tr}\{\rho(0)\, B^2\} \leq 
\norm{B^2}\, \mbox{Tr}\{\rho(0)\} =\norm{B}^2 \ ,
\label{a210}
\end{eqnarray}
where $\norm{C}$ indicates the operator 
norm of an arbitrary Hermitian operator
$C$ (largest eigenvalue in modulus).
From (\ref{a60}) we can infer that the eigenvalues 
and hence the operator norm of $A$ and $B$ are equal.
Altogether, we thus can rewrite (\ref{a200}) as
\begin{eqnarray}
|R_1| & \leq & \norm{A}\sqrt{S}
\ ,
\label{a220}
\\
S & := & \mbox{Tr}\{\rho(0)\, Q^2\} \ .
\label{a230}
\end{eqnarray}
Evaluating the trace in (\ref{a230}) by 
means of the energy basis $|n\rangle$ 
and exploiting (\ref{a90}) yields
\begin{eqnarray}
S=\sum\langle n|\rho(0)\, Q^2|n\rangle
=\sum_n \rho_{nn}(0) (1-g_n)^2 \ .
\label{a240}
\end{eqnarray}
One readily verifies that 
$(1-x)^2\leq |1-x^2|$ for any $x\geq 0$.
Recalling that $\rho_{nn}(0)=p_n > 0$
thus implies
\begin{eqnarray}
\rho_{nn}(0) (1-g_n)^2\leq p_n |1-g_n^2|=|p_n-p_n g^2_n| \ .
\label{a250}
\end{eqnarray}
Since $p_n g^2_n=\tilde p_n$, see (\ref{a70}),
we finally can rewrite (\ref{a240}) as
\begin{eqnarray}
S\leq\sum_n |p_n-\tilde p_n| \ .
\label{a260}
\end{eqnarray}

The treatment of $R_2$ in (\ref{a150}) is 
similar and thus only briefly sketched:
\begin{eqnarray}
|R_2|^2 & = & |\mbox{Tr}\{(B P\sigma) (\sigma Q)\}|^2
\nonumber
\\
& \leq & \mbox{Tr}\{B P\rho(0) P B\}\, \mbox{Tr}\{Q\rho(0)Q\}
\nonumber
\\
& = & \mbox{Tr}\{B^2 \tilde\rho(0)\}\, \mbox{Tr}\{Q^2\rho(0)\}
\leq \norm{A}^2 S \ .
\label{a270}
\end{eqnarray}

Introducing (\ref{a220}), (\ref{a260}), and (\ref{a270})
into (\ref{a150}) yields
\begin{eqnarray}
|\Delta|\leq |R_1|+|R_2|\leq 2\norm{A}\sqrt{\sum_n |p_n-\tilde p_n|}
\ .
\label{a280}
\end{eqnarray}
Obviously, $\Delta$ in (\ref{a150}) remains 
unchanged when adding an arbitrary real 
constant $c$ to $A$. 
Hence, the inequality (\ref{a280}) with $\norm{A+c}$
instead of $\norm{A}$ on the right 
hand side remains valid for 
arbitrary $c$.
The minimum over all $c$ is assumed
when the largest and smallest eigenvalues of
$A+c$ are of opposite sign and equal modulus,
yielding
\begin{eqnarray}
|\Delta|\leq \da \sqrt{\sum_n |p_n-\tilde p_n|} \ ,
\label{a290}
\end{eqnarray}
where $\da$ is the difference
between the largest and smallest 
eigenvalues of $A$.
Recalling the definition of $\Delta$
in (\ref{a150}), we recover the
announced result (\ref{a10}).

So far we have assumed that $\rho_{nn}(0)>0$ 
for all $n$, see above (\ref{a70}).
More generally, we may define
\begin{eqnarray}
\rho^\epsilon(0)
& := &
\frac{1}{1+b\epsilon}
\left(
\rho(0)+\epsilon \sum_n \frac{1}{n^2}|n\rangle\langle n|
\right)
\ ,
\label{a300}
\\
b
& := &
\sum_n \frac{1}{n^2} \ .
\label{a310}
\end{eqnarray}
One readily confirms that 
$0 < b \leq \sum_{k=1}^\infty k^{-2}=\pi^2/6$
and that $\rho^\epsilon(0)$ is Hermitian, non-negative,
and of unit trace, i.e., a well defined 
density operator for any $\epsilon\geq 0$.
Analogously, all quantities deriving from
$\rho(0)$ now become $\epsilon$ dependent 
and acquire an additional index $\epsilon$,
in particular $p^\epsilon_n:=\rho^\epsilon_{nn}(0)$
and $\tilde p^\epsilon_n:=\tilde \rho^\epsilon_{nn}(0)$,
while the preset $\tilde p_n$ 
(without index $\epsilon$)
are kept fixed.
The case of actual interest is thus 
recovered in the limit $\epsilon\to 0$.
Moreover, one can show that the
off-diagonal matrix elements 
$\tilde\rho^\epsilon_{mn}(0)$
identically vanish if either 
$p_m=0$ or $p_n=0$
and that all other matrix elements
$\tilde\rho^\epsilon_{mn}(0)$
converge towards a finite limit 
for $\epsilon\to 0$.
In particular, $\tilde p_n^\epsilon$ 
approaches $\tilde p_n$.
As a consequence, also $\tilde\rho^\epsilon(0)$
itself approaches a well defined 
limit, which exhibits the preset level
populations $\tilde p_n$.

For any $\epsilon>0$ one can infer from
(\ref{a300}) that 
$p^\epsilon_n:=\rho^\epsilon_{nn}(0)>0$ 
for all $n$, i.e., the result (\ref{a10}) 
is valid.
For continuity reasons, the same result must
still remain valid in the limit
$\epsilon\to 0$, in which we are actually 
interested.

\section*{Appendix B}
This appendix substantiates the statement 
above (\ref{203}) in the main text.
Before actually recalling this statement
itself, it is necessary to recall the  
setup and the notation:
We consider two arbitrary density 
operators $\rho(t)$ and $\tilde\rho(t)$
with identical initial conditions,
\begin{eqnarray}
\rho(0)=\tilde\rho(0) \ ,
\label{b20}
\end{eqnarray}
but whose time evolution is
governed by different Hamiltonians,
namely $H$ from (\ref{5}) and
$\tilde H$ from (\ref{204}),
respectively.
In other words, the eigenvectors
$|n\rangle$ of $H$ and $\tilde H$
must be identical, while the
eigenvalues $E_n$ and $\tilde E_n$ 
may be different.
The main goal of this appendix is to 
show that
\begin{eqnarray}
|\langle A\rangle_{\!\rho(t)} 
- \langle A\rangle_{\!\tilde\rho(t)}|
\leq |t| \, \da \, 
\max\limits_{n\in I}|\tilde E_n-E_n|/\hbar
\label{b30}
\end{eqnarray}
for arbitrary $t$ and $A$,
where $\da$ is the difference between the 
largest and smallest eigenvalues of $A$,
and $I$ is defined in (\ref{203a}).
Similarly as below (\ref{a10}) one 
sees that this amounts to a detailed 
quantitative justification of the 
statement above (\ref{203})
in the main text.

Note that there is a slight 
notational difference between 
the main text and this appendix:
In the main text, one starts out with
$\tilde \rho(t)$, whose dynamics is
governed by $H$, and whose
level populations 
$\tilde p_n:=\tilde\rho_{nn}(0)$
define the set $I$ via (\ref{203a}).
Then the Hamiltonian $H$ is 
replaced by $\tilde H$, but
for notational convenience the
modified density operator is 
still named $\tilde\rho(t)$.
In the present appendix, the
two density operators carry
the two different names
$\rho(t)$ and $\tilde\rho(t)$,
respectively.
Due to (\ref{b20}), their initial 
level populations are identical, 
i.e., we have
\begin{eqnarray}
p_n:=\rho_{nn}(0)=\tilde p_n:=\tilde\rho_{nn}(0)
\label{b40}
\end{eqnarray}
throughout this appendix
(but not in the main text).
Accordingly, (\ref{203a}) can 
be rewritten as 
\begin{eqnarray}
I = \{n \, |\, p_n>0\,\} \ .
\label{b50}
\end{eqnarray}

Similarly as in (\ref{a20})-(\ref{a45}) 
one finds for $\tilde\rho(t)$ as specified 
below (\ref{b20}) that
\begin{eqnarray}
\langle A\rangle_{\!\tilde\rho(t)} 
& = & 
\mbox{Tr}\{\tilde\rho(t)A\}
\ ,
\label{b90}
\\
\tilde\rho(t) 
& = & 
\tilde \propa_t \rho(0) \tilde \propa_t^\dagger
\ ,
\label{b100}
\\
\tilde \propa_t 
& := & 
e^{-i\tilde Ht/\hh} = \propa_t'\propa_t
\ ,
\label{b110}
\\
\propa_t' 
& := &
e^{i(H-\tilde H)t/\hh} \ .
\label{b120}
\end{eqnarray}
The last identity in (\ref{b110})
relies on the fact that $H$ and 
$\tilde H$ commute.
Together with (\ref{a20})-(\ref{a45}) 
it follows that $\tilde\rho(t)=\propa_t'\rho(t)(\propa_t')^\dagger$
and due the cyclic invariance of the trace that
\begin{eqnarray}
\Delta 
& := &  
\langle A\rangle_{\!\rho(t)} 
- \langle A\rangle_{\!\tilde\rho(t)} 
=
\mbox{Tr}\{\rho(t)\,B\}
\ ,
\label{b130}
\\
B & := &
A-(\propa_t')^\dagger A \propa_t'\ .
\label{b140}
\end{eqnarray}

Evaluating the trace in (\ref{b130})
by means of the  eigenbasis of 
$\rho(t)$ yields 
\begin{eqnarray}
|\Delta| & \leq & \max_{\norm{\psi}=1} |\Delta_{\psi}|
\ ,
\label{b150}
\\
\Delta_\psi & := & \langle\psi|B|\psi\rangle \ ,
\label{b160}
\end{eqnarray}
where the maximization in (\ref{b150}) is
over all normalized vectors $|\psi\rangle$.

For an arbitrary but fixed 
vector $|\psi\rangle$ of unit norm
we can rewrite (\ref{b160}) with (\ref{b140}) as
\begin{eqnarray}
\Delta_\psi 
& = & 
\langle\psi|A|\psi\rangle -
\langle\psi'|A|\psi'\rangle
\ ,
\label{b170}
\\
|\psi'\rangle 
& := & \propa_t'|\psi\rangle \ .
\label{b180}
\end{eqnarray}
With the definition 
\begin{eqnarray}
|\chi\rangle := |\psi'\rangle-|\psi\rangle
\label{b190}
\end{eqnarray}
we can conclude that
\begin{eqnarray}
\langle\psi'|A|\psi'\rangle
& = &
\langle\psi'|A|\psi\rangle + d_1
\ ,
\label{b200}
\\
d_1 
& := &
\langle\psi'|A|\chi\rangle
\ ,
\label{b210}
\\
\langle\psi'|A|\psi\rangle
& = &
\langle\psi|A|\psi\rangle + d_2
\ ,
\label{b220}
\\
d_2
& := &
\langle\chi|A|\psi\rangle \ .
\label{b230}
\end{eqnarray}
By means of (\ref{b200}) and (\ref{b220}) 
we can conclude with (\ref{b170}) that
\begin{eqnarray}
|\Delta_\psi| \leq |d_1| + |d_2| \ .
\label{b240}
\end{eqnarray}
From the definition (\ref{b210}) and the
Cauchy-Schwarz inequality it follows that
\begin{eqnarray}
|d_1|^2=|\langle\chi| (A|\psi'\rangle)|^2
\leq
\langle\chi|\chi\rangle \langle\psi'|A^2|\psi'\rangle \ .
\label{b250}
\end{eqnarray}
Since we assumed that $|\psi\rangle$ is normalized,
also $|\psi'\rangle$ in (\ref{b180})
will be normalized and the last factor 
in (\ref{b250}) can be upper bounded by
$\norm{A^2}=\norm{A}^2$,
where $\norm{A}$ is the operator norm of $A$
(see also below (\ref{a210})).
Exactly the same upper bound can be
obtained for $d_2$ in (\ref{b230}).
With (\ref{b240}) we thus arrive at
\begin{eqnarray}
|\Delta_\psi| \leq 2 \norm{A}\sqrt{\langle\chi|\chi\rangle}
\ .
\label{b260}
\end{eqnarray}
Obviously, $\Delta$ in (\ref{b130}) remains 
unchanged when adding an arbitrary real constant 
$c$ to $A$.
Exactly as below (\ref{a280}) one thus can conclude that
\begin{eqnarray}
|\Delta_\psi| \leq  \da \sqrt{\langle\chi|\chi\rangle} \ .
\label{b270}
\end{eqnarray}

Rewriting $|\psi\rangle$ as $\sum_n c_n\,|n\rangle$
with $c_n:=\langle n|\psi\rangle$,
the normalization takes the form 
$\sum_n |c_n|^2=1$.
Furthermore, we can infer from
(\ref{5}), (\ref{204}), (\ref{b120}), 
and (\ref{b180}) that
\begin{eqnarray}
|\psi'\rangle & = & \sum_n e^{i a_n} c_n |n\rangle
\ ,
\label{b280}
\\
a_n & := & (E_n-\tilde E_n)t/\hh
\label{b290}
\end{eqnarray}
and from (\ref{b190}) that
\begin{eqnarray}
\langle\chi|\chi\rangle
=
\sum_n |c_n-e^{ia_n}c_n|^2
=
\sum_n |c_n|^2 |1-e^{ia_n}|^2
. 
\label{b300}
\end{eqnarray}
One readily verifies that 
$|1-e^{ia}|=2|\sin(a/2)|\leq |a|$
for arbitrary $a\in \RR$,
yielding
\begin{eqnarray}
\langle\chi|\chi\rangle
\leq
\sum |c_n|^2 \,|a_n|^2\leq 
\max_n |a_n|^2
\ .
\label{b310}
\end{eqnarray}
By introducing (\ref{b310}) into (\ref{b270}) 
we can conclude
\begin{eqnarray}
|\Delta_\psi| \leq  \da \max_n |a_n|
\ .
\label{b320}
\end{eqnarray}
Since this bound is independent of $|\psi\rangle$,
one finds by means of (\ref{b290}), (\ref{b150}),
and (\ref{b130}) that
\begin{eqnarray}
|\langle A\rangle_{\!\rho(t)} 
- \langle A\rangle_{\!\tilde\rho(t)}|
\leq |t| \, \da \, 
\max\limits_{n}|\tilde E_n-E_n|/\hbar
\ .
\label{b330}
\end{eqnarray}

Exploiting the definition
$\rho_{mn}(0):=\langle m|\rho(0)|n\rangle$
(see below (\ref{10}))
and the Cauchy-Schwarz 
inequality one can readily show that 
$|\rho_{mn}(0)|^2 \leq 
\rho_{mm}(0)\rho_{nn}(0)=p_m\, p_n$
(see also (\ref{b40})).
It follows that only those summands
in (\ref{10}) are non-zero for which
both $m$ and $n$ are contained in the
set $I$ from (\ref{b50}).
Without loss of generality we thus can
focus on the case that $\tilde E_n=E_n$
for all $n\not\in I$.
As a consequence, it is sufficient
to maximize in (\ref{b330}) over all
$n\in I$, i.e., we recover the announced 
final result (\ref{b30}).

\section*{Appendix C}
The purpose of this appendix is to 
show that
(\ref{375}) with $G(t)$ from 
(\ref{360}) and (\ref{450})
is fulfilled in very good 
approximation if $\rho(0)$ 
is a mixed state
of low purity, that is, if
\begin{eqnarray}
P:=\pu \ll 1
\ .
\label{d5}
\end{eqnarray}

Conceptually, the subsequent considerations are
somewhat similar to the explorations of dynamical
typicality in  \cite{bar09,fin09,mul11,rei18}.
Technically, the calculations are particularly close 
to those in  \cite{rei18}.

To begin with, we denote the eigenvalues and eigenvectors of
$\rho(0)$ by $r_n$ and $|\varphi_n\rangle$, 
respectively, implying
\begin{eqnarray}
\rho(0) & = & \sum_n r_n \, |\varphi_n\rangle\langle\varphi_n|
\ ,
\label{d11}
\\
P  & = &\sum_n r_n^2\ll 1
\ ,
\label{d10}
\end{eqnarray}
where $r_n\geq 0$ and $\sum_n r_n=1$.
Next, we consider an ensemble of 
(not necessarily normalized)
random vectors $|\varphi\rangle$,
defined via
\begin{eqnarray}
|\varphi\rangle=
\sum_n c_n\, \sqrt{r_n}\, |\varphi_n\rangle
\ ,
\label{d13}
\end{eqnarray}
where the real and imaginary parts 
of the $c_n$'s are independent, 
Gaussian distributed random variables 
of mean zero and variance $1/2$.
Indicating averages over the $c_n$'s
by the symbol  $[...]_c$, one readily 
confirms that
\begin{eqnarray}
\left[c^\ast_m c_n\right]_c 
& = & 
\delta_{mn}
\ , 
\label{d13a}
\\
\left[c^\ast_j c_k c^\ast_m c_n\right]_c
& = &
\delta_{jk}\delta_{mn}+\delta_{jn}\delta_{km}
\label{d13b}
\end{eqnarray}
for arbitrary indices $m,n,j,k$.
Given any Hermitian operator $B$,
it then follows from (\ref{d13})-(\ref{d13b}) 
by means of a straightforward calculation
(see also  \cite{rei18})  that
\begin{eqnarray}
\mu_B & := & \left[\langle\varphi|B|\varphi\rangle\right]_c
= \mbox{Tr}\{\rho(0)  B\}
\ ,
\label{d16}
\\
\sigma_B^2 & := &
\left[
\left(\langle\varphi|B|\varphi\rangle
- \mu_B
\right)^2
\right]_c
=
\mbox{Tr}\{[\rho(0)  B]^2\}
\ .
\label{d17}
\end{eqnarray}
By analogous arguments as above 
 (\ref{a190}), one can deduce 
from (\ref{d17}) that
$\sigma_B^2  \leq \mbox{Tr}\{B^2\, [\rho(0)]^2\}$.
Evaluating the trace by means of the
eigenbasis of $B$ and exploiting the
definition of the purity $P$ in (\ref{d5}) 
then yields
\begin{eqnarray}
\sigma_B^2  \leq \norm{B}^2\, P
\ .
\label{d18}
\end{eqnarray}

Choosing $B=\id$,
it follows from (\ref{d10}), (\ref{d16}), 
and (\ref{d18}) that
$[\langle\varphi|\varphi\rangle]_c = 1$ and 
$[\left(\langle\varphi|\varphi\rangle- 1 \right)^2]_c\leq P$.
Invoking the Chebyshev inequality from 
probability theory, one thus can conclude that
\begin{eqnarray}
& & 
\pr\!\left(
| \langle\varphi|\varphi\rangle - 1 |\leq P^{\frac{1}{3}}
\right)
\geq 1-P^{\frac{1}{3}} \ ,
\label{d19}
\end{eqnarray}
where the left hand side denotes the probability
that $|\langle\varphi|\varphi\rangle -1 |\leq P^{\frac{1}{3}}$
when randomly sampling vectors $|\varphi\rangle$ 
according to (\ref{d13}).
Due to (\ref{d5}), the vast majority of all 
vectors $|\varphi\rangle$ in (\ref{d13}) thus have 
norms very close to unity.

Choosing $B$ as in (\ref{a60}),
it follows from (\ref{a50}) and (\ref{d16})
that
\begin{eqnarray}
\left[ \langle A\rangle_{\!\varphi(t)}\right]_c 
=
\langle A\rangle_{\!\rho(t)} 
\ ,
\label{d20}
\end{eqnarray}
where we have introduced
\begin{eqnarray}
\langle A\rangle_{\!\varphi(t)}
& := &
\langle\varphi(t)|A|\varphi(t)\rangle = \mbox{Tr}\{\rho_{\varphi(t)} A\}
\ ,
\label{d21}
\\
\rho_\varphi(t) 
& := &
|\varphi(t)\rangle\langle\varphi(t)|
\ ,
\label{d22}
\\
|\varphi(t)\rangle
& := &
\propa_t |\varphi\rangle
\ ,
\label{d22a}
\end{eqnarray}
and where the propagator $\propa_t$ is 
defined in equation (\ref{6}).
Observing that the operator norm of $B$ in (\ref{a60})
is identical to the operator norm of $A$, we can infer
from (\ref{d18}) and the above definitions that
\begin{eqnarray}
\left[ \{
\langle A\rangle_{\!\varphi(t)}  - \langle A\rangle_{\!\rho(t)} 
\}^2\right]_c 
\leq \norm{A}\, P
\ .
\label{d24}
\end{eqnarray}
Similarly as below (\ref{a280}), the operator norm 
$\norm{A}$ on the right hand side of (\ref{d24}) can 
furthermore be replaced
by $\da/2$, where $\da$ is the 
measurement range of the observable $A$ 
(largest minus smallest eigenvalue).
Invoking Chebyshev's inequality once more, 
one thus arrives at
\begin{eqnarray}
\pr\!\left(
| \langle A\rangle_{\!\varphi(t)}  - \langle A\rangle_{\!\rho(t)}  |\leq \frac{\da}{2}P^{\frac{1}{3}}
\right)
\geq 1- P^{\frac{1}{3}}. 
\label{d25}
\end{eqnarray}
In view of (\ref{d5}), the vast
majority of all vectors $|\varphi\rangle$ 
in (\ref{d13}) thus exhibit expectation 
values $\langle A\rangle_{\!\varphi(t)}$, whose
deviations from the ensemble 
average $\langle A\rangle_{\!\rho(t)} $ 
are very small compared 
to full range $\da$ over which those expectation 
values in principle could vary.

Recalling the definition of $\rhobar$  in (\ref{15}) 
and defining in the same vein the auxiliary observable 
\begin{eqnarray}
\Adia:=\sum_n A_{nn}\, |n\rangle \langle n|
\ ,
\label{c50}
\end{eqnarray}
one can conclude that
\begin{eqnarray}
\langle A\rangle_{\!\rhobar }
=
\mbox{Tr}\{\rho(0)\Adia\}
 \ .
\label{d32}
\end{eqnarray}
The corresponding quantities for the pure state
$\rho_{\varphi}(t)$ in (\ref{d22}) are defined as
\begin{eqnarray}
\langle A\rangle_{\!\varphi}^{\dia}
& := & \mbox{Tr}\{\rho_{\!\varphi}^{\dia} A\}
\label{d33}
\\
\rho_{\!\varphi}^{\dia}
& := &
\sum_n \langle n|\rho_{\!\varphi}(0)|n\rangle \, |n\rangle\langle n|
\ .
\label{d34}
\end{eqnarray}
Similarly as in (\ref{d32}), it readily follows that
\begin{eqnarray}
\langle A\rangle_{\!\varphi}^{\dia}
=
\mbox{Tr}\{\rho_{\!\varphi}(0) \Adia\}
=\langle\varphi |\Adia|\varphi\rangle
 \ ,
\label{d35}
\end{eqnarray}
where we exploited (\ref{d22}) and (\ref{d22a}) in the last step.
Upon choosing $B=\Adia$ one then finds along the very
same line of reasoning as in (\ref{d20})-(\ref{d24}) that
\begin{eqnarray}
\left[ \{
\langle A\rangle_{\!\varphi}^{\dia} - \langle A\rangle_{\!\rhobar}
\}^2\right]_c 
\leq \norm{\Adia}\, P
\ ,
\label{d35a}
\end{eqnarray}
and by observing that $\norm{\Adia}\leq \norm{A}$
it follows as in (\ref{d25}) that
\begin{eqnarray}
\pr\!\left(
| \langle A\rangle_{\!\varphi}^{\dia} - \langle A\rangle_{\!\rhobar}  |\leq \frac{\da}{2}P^{\frac{1}{3}}
\right)
\geq 1- P^{\frac{1}{3}} 
\, . \ \ \ 
\label{d36}
\end{eqnarray}

Similarly as in (\ref{460}), we define for the pure state
$|\varphi(t)\rangle$ in (\ref{d22a}) the overlap
\begin{eqnarray}
\chi_{\varphi}(t) := \langle\varphi(t)|\varphi(0)\rangle
\ .
\label{d37}
\end{eqnarray}
The corresponding quantity for the mixed state
$\rho(0)$ in (\ref{d11}) is defined as
\begin{eqnarray}
\chi(t) & := & \mbox{Tr}\{\rho(0)\, \propa_t^\dagger\}
\ .
\label{d37a}
\end{eqnarray}
In view of (\ref{6}), this definition is
equivalent to  (\ref{450}) in the 
main text.
By exploiting (\ref{d13}) and (\ref{d22a})
one can rewrite (\ref{d37}) as
\begin{eqnarray}
\chi_{\varphi}(t) & = & 
\sum_{mn}c_m^\ast c_n\sqrt{r_mr_n} \langle\varphi_m| \tilde\varphi_n\rangle
\ ,
\label{d38}
\\
| \tilde\varphi_n\rangle
& := & \propa_t^\dagger|\varphi_n\rangle
\ .
\label{d38b}
\end{eqnarray}
Together with (\ref{d13}), (\ref{d13a}), and (\ref{d37}),
a straightforward calculation then yields the result
\begin{eqnarray}
\left[\chi_{\varphi}(t) \right]_c
= 
\chi(t)
\ .
\label{d39}
\end{eqnarray}
Likewise, the variance
\begin{eqnarray}
\sigma_\chi^2:=\left[ \left| \chi_{\varphi}(t) - \chi(t)\right|^2\right]_c
\label{d40}
\end{eqnarray}
can be evaluated with the help of (\ref{d13b}) to yield
\begin{eqnarray}
\sigma^2_\chi = 
\sum_{mn} r_mr_n\,| \langle\varphi_m| \tilde\varphi_n\rangle |^2
\ .
\label{d41}
\end{eqnarray}
Rewriting the summands on the right hand side of
(\ref{d41}) as $v_{mn}w_{mn}$ with
$v_{mn}:= r_m | \langle\varphi_m|\tilde \varphi_n\rangle|$,
$w_{mn}:= r_n | \langle\varphi_m|\tilde \varphi_n\rangle|$,
and invoking the Cauchy-Schwarz inequality, one can 
conclude that
\begin{eqnarray}
(\sigma^2_\chi)^2 
& \leq & V\, W
\ ,
\label{d42}
\\
V & := & \sum_{mn}v_{mn}^2
\ ,
\label{d42a}
\\
W & := & \sum_{mn}w_{mn}^2
\ .
\label{d42b}
\end{eqnarray}
It follows that
\begin{eqnarray}
V & = &
\sum_{mn} r_m^2 \,| \langle\varphi_m|\tilde\varphi_n\rangle |^2
\nonumber
\\
& = &
\sum_{m} r_m^2 \sum_n \langle\varphi_m|\tilde\varphi_n\rangle
\langle\tilde \varphi_n|\varphi_m\rangle
=
\sum_m r_m^2 
\ .
\label{d43}
\end{eqnarray}
The same result is readily recovered also for $W$ from (\ref{d42b}).
With (\ref{d10}) we thus can conclude that $\sigma^2_\chi\leq P$.
Due to (\ref{d40}) and Chebyshev's inequality it follows that
\begin{eqnarray}
\pr\!\left(  \left |\chi_{\varphi}(t) - \chi(t)\right | \leq P^{\frac{1}{3}} \right) \geq 1- P^{\frac{1}{3}}
\ .
\label{d45}
\end{eqnarray}

Finally, by similar arguments as above one can also
show that the purity of the diagonal ensemble 
from (\ref{d34}),
\begin{eqnarray}
P_{\!\varphi} := \mbox{Tr}\left\{(\rho_\varphi^{\dia})^2\right\} 
\ ,
\label{d46}
\end{eqnarray}
satisfies the relation
\begin{eqnarray}
\left[ P_{\!\varphi} \right]_c \leq 2\, P
\ .
\label{d47}
\end{eqnarray}
Moreover, one readily infers from (\ref{d34}) that
\begin{eqnarray}
\max_{n} \langle n|\rho_{\!\varphi}(0)|n\rangle \leq P_{\!\varphi}
\ .
\label{d48}
\end{eqnarray}
Since the left hand side of (\ref{d48}) is non-negative,
one can apply Markov's inequality to conclude 
\begin{eqnarray}
\pr\!\left( \max_{n}\, \langle n|\rho_{\!\varphi}(0)|n\rangle\leq P^{\frac{1}{2}}\right) 
\geq 1- 2P^{\frac{1}{2}}
\ .
\label{d49}
\end{eqnarray}

So far, the random vectors $|\varphi\rangle$ 
in (\ref{d13}) are in general not 
normalized. But, as mentioned below
(\ref{d19}), the vast majority 
among them is almost of unit 
length. 
Hence, if we replace every given
$|\varphi\rangle$ in (\ref{d13}) by 
its strictly normalized counterpart
\begin{eqnarray}
|\psi\rangle
:= 
\frac{|\varphi\rangle}{\sqrt{\langle\varphi|\varphi\rangle}}
\ ,
\label{d50}
\end{eqnarray}
then the ``new'' expectation values
$\langle\psi|B|\psi\rangle$
will mostly remain very close to the
``old'' ones, i.e., to
$\langle\varphi|B|\varphi\rangle$
for any given Hermitian operator $B$.
Essentially, this is a consequence of the
relation
\begin{eqnarray}
\langle\psi|B|\psi\rangle
=
\frac{\langle\varphi|B|\varphi\rangle}{\langle\varphi |\varphi\rangle}
\ ,
\label{d51}
\end{eqnarray}
which follows from (\ref{d50}),
and of the fact that $\langle\varphi |\varphi\rangle$ 
is very close to unity for most $|\varphi\rangle$'s 
according to (\ref{d5}) and (\ref{d19}).
Defining quantities analogous to those in
(\ref{d21})-(\ref{d22a}) for $\psi$ 
instead of $\varphi$,
it follows with (\ref{d5}) and (\ref{d25}) that the vast majority
of the normalized random vectors $|\psi\rangle$
in (\ref{d50}) still satisfy
in very good approximation the relation
\begin{eqnarray}
\langle A\rangle_{\!\psi(t)}  = \langle A\rangle_{\!\rho(t)}  
\ .
\label{d52}
\end{eqnarray}
Likewise, with analogous definitions as in 
(\ref{d33}), (\ref{d34}) for $\psi$ 
instead of $\varphi$, one 
can conclude from (\ref{d5}) and (\ref{d36}) 
that most $|\psi\rangle$'s will satisfy
\begin{eqnarray}
\langle A\rangle_{\!\psi}^{\dia} = \langle A\rangle_{\!\rhobar} 
\label{d53}
\end{eqnarray}
in very good approximation.
Finally, defining $\chi_\psi(t)$ analogously as in (\ref{d37}),
one sees, similarly as in (\ref{d51}),
that $\chi_\psi(t)$
is equal to
$\chi_\varphi(t)/\langle\varphi |\varphi\rangle$
and that
$\langle n|\rho_{\!\psi}(0)|n\rangle$ is equal to
$\langle n|\rho_{\!\varphi}(0)|n\rangle/\langle\varphi |\varphi\rangle$.
Together with (\ref{d5}), (\ref{d19}), (\ref{d45}), 
and (\ref{d49}), one thus can conclude as before 
that the relations
\begin{eqnarray}
& & 
\chi_{\psi}(t) = \chi(t)
\ ,
\label{d54}
\\
& & 
\max_{n} \langle n|\rho_{\!\psi}(0)|n\rangle\ll 1
\label{d55}
\end{eqnarray}
will be satisfied in very good approximation for
most $|\psi\rangle$'s.
A more detailed quantitative demonstration
that all four approximations (\ref{d52})-(\ref{d55}) 
will be {\em simultaneously} 
fulfilled very well by most $|\psi\rangle$'s
can be worked out analogously as in 
 \cite{rei18}.

At this point, a subtle notational difference between
the main text and this appendix comes into play:
In the main text, the result (\ref{375}) with
$G(t)$ from (\ref{360}) was derived under the condition
that $\rho(0)$ is a pure state, see (\ref{220}), and hence
$G(t)$ can be written in the form (\ref{470}).
In the present appendix, $\rho(0)$ represents 
a mixed state of low purity according to (\ref{d5}).
In turn, the above mentioned result for pure states
in the main text should now be rewritten for the 
pure states $|\psi(t)\rangle$ considered in this 
appendix as
\begin{eqnarray}
\langle A\rangle_{\!\psi(t)} 
& = & 
\langle A\rangle_{\!\psi}^{\dia}
+
|\chi_\psi(t)|^2\, 
\left[
\langle A\rangle_{\!\psi(0)} 
- \langle A\rangle_{\!\psi}^{\dia}
\right] \, . \ \ \
\label{d56}
\end{eqnarray}
One the other hand, since $\rho(0)$ in this appendix
is a mixed state of low purity according to (\ref{d5}),
we know that most $|\psi\rangle$'s simultaneously
fulfill (\ref{d52})-(\ref{d55}) in very good approximation.
If we choose one of those $|\psi\rangle$'s in (\ref{d56}),
we obtain with (\ref{d52})-(\ref{d55}) in very good 
approximation the result
\begin{eqnarray}
\langle A\rangle_{\!\rho(t)} 
& = & 
\langle A\rangle_{\!\rhobar}
+
|\chi(t)|^2\, \left[\langle A\rangle_{\!\rho(0)} 
- \langle A\rangle_{\!\rhobar}\right]
 \, , \ \ \
\label{d57}
\end{eqnarray}
with $\chi(t)$ from  (\ref{d37a}).
Since the latter equation is equivalent to  (\ref{450}),
we thus have proven that (\ref{375}) in the main
text in fact also holds true for mixed
states $\rho(0)$ of low purity, as announced 
at the beginning of this appendix.

In the above conclusion, we have tacitly taken for
granted one more assumption, namely that there
exists at least one $|\psi\rangle$ which satisfies
(\ref{d52})-(\ref{d55}) very well, and which at the
same time satisfies the preconditions for (\ref{d56}),
as discussed in sections \ref{s3} and \ref{s4}.
While a rigorous justification of this extra 
assumption seems to be a quite daunting task,
it also seems quite reasonable to expect that
the assumption will be fulfilled if (and only if) 
the mixed state $\rho(0)$ itself satisfies those 
preconditions from sections \ref{s3} and \ref{s4}.

Finally, we turn to the case that the mixed
state $\rho(0)$ is {\em not} of low purity 
(but still not a pure state).
In such a case, there is no reason to expect
that (\ref{d52})-(\ref{d56}) will be simultaneously
fulfilled for {\em most} $|\psi\rangle$'s.
However, one may still expect that (\ref{d52})-(\ref{d56}) 
will be simultaneously fulfilled for at least {\em one} 
$|\psi\rangle$, at least for {\em some}
such $\rho(0)$'s.
If so, (\ref{d57}) and thus (\ref{375}) in the main text
still remain true even when the purity of 
$\rho(0)$ is not small.


\end{document}